\documentclass[aps,prd,12pt,preprint,tightenlines,superscriptaddress,
amsfonts,byrevtex,showpacs]{revtex4}

\usepackage{amssymb}
\usepackage{amsmath}
\usepackage{latexsym}
\usepackage{graphicx}
\usepackage{color}

\newtheorem{prop}{Proposition}
\newcommand{\DEM}{\noindent {\bf Proof.}\ \ }
\newcommand{\QED}{\ \hfill \rule{0.5em}{0.5em} }

\def\setR{\mathbb{R}}
\def\setN{\mathbb{N}}
\def\setC{\mathbb{C}}
\def\K{\mathcal{K}}
\def\1{\mathbf {Id} }
\def\z {{\cal Z}}

\def\dz {\frac{d}{d{\cal Z}}\,}

\def\bc{\bar\partial_{\alpha}}
\def\bpc{\bar\partial'_{\beta'}}
\def\ab{\frac{\partial x^{\alpha}}{\partial X^a}\frac{\partial x'^{\beta'}}{\partial X'^{b'}}\;}
\def\a{\frac{\partial x^{\alpha}}{\partial X^a}\;}
\def\bb{\frac{\partial x'^{\beta'}}{\partial X'^{b'}}\;}
\makeatletter
\renewcommand\thesection{\@Roman\c@section}
\renewcommand\theequation{\@arabic\c@section.\@arabic\c@equation}

\renewcommand{\theequation}{\thesection.\arabic{equation}}
\makeatother

\begin{document}

\title{``Massless'' vector field in de Sitter Universe
}

\author{
T. Garidi$^{1}$, J-P. Gazeau$^{1}$, S. Rouhani$^2$ and M.V. Takook$^{3}$\\
{\it $1$ - Boite 7020, APC, CNRS UMR 7164, Universit\'e Paris~7,
Denis Diderot,
F-75251 Paris Cedex 05, France.\\
$2$ - Plasma Physics Research Center, Islamic Azad
University,P.O.BOX 14835-157, Tehran, IRAN\\
$3$ - Department of Physics, Razi University, Kermanshah, Iran. }}
 \email{garidi@ccr.jussieu.fr,
gazeau@ccr.jussieu.fr, takook@razi.ac.ir}

\date{\today}

\begin{abstract}

In the present work the massless vector field in the de Sitter (dS)
space has been quantized. ``Massless" is used here by reference to
conformal invariance and propagation on the dS light-cone whereas
``massive" refers to those dS fields which contract at zero
curvature unambiguously to massive fields in Minkowski space. Due to
the gauge invariance of the massless vector field, its covariant
quantization requires  an indecomposable representation of the de
Sitter group and an indefinite metric quantization. We will work
with a specific gauge fixing which leads to the simplest one among
all possible related Gupta-Bleuler structures. The field operator
will be defined with the help of coordinate independent de Sitter
waves (the modes) which are simple to manipulate and most adapted to
group theoretical matters. The physical states characterized by the
divergencelessness condition will for instance be easy to identify.
The whole construction is based on analyticity requirements in the
complexified pseudo-Riemanian manifold for the modes and the
two-point function.

\end{abstract}
\maketitle

\newpage
\section{Introduction}

In a previous work, the so-called ``massive'' vector field in dS
space has been considered \cite{gata}. Unitary irreducible
representation (UIR) in the principal series of the  de Sitter group
$SO_0(1,4)$, with  Casimir operator eigenvalue
$<Q_\nu^{(1)}>=\nu^2+\frac{1}{4},\;\;\nu \geq 0$ and the
corresponding ``mass''$m_p^2=H^2(\nu^2+\frac{1}{4})$, are associated
with that vector field. The  interpretation in term of mass of this
field is made possible by carrying out the null curvature limit.
Indeed, the principal series of  UIR's admits a massive Poincar\'e
group UIR in the limit $H=0$ \cite{eric,mini}. However, there is
another vector UIR of the de Sitter group with a minkowskian limit.
In other words it is the UIR having  a natural extension to the
conformal group SO$_0(2,4)$, which is equivalent to the massless
spin $1$ UIR of the conformal extension  of the Poincar\'{e} group
\cite{barutbohm,anflafrons}.  The corresponding field obeys a
conformal invariant field equation and the minkowskian
interpretation is that of a massless field. This UIR belongs to the
discrete series of UIR of the dS group corresponding to eigenvalue
$<Q_\nu^{(1)}>=0$  of the Casimir operator, and this value
characterizes the field  we call ``massless'' vector field.

The covariant quantization of the massless vector field raises
various problems, analogous to those encountered in the
quantization of the electromagnetic field in Minkowski space.
First of all one should note that the field equation admits gauge
solutions. Therefore one is free to use  a gauge fixing parameter
$c$. Now it is known \cite{binegar,ga} that the quantization of gauge
invariant theories usually requires
quantization \emph{\`a la}  Gupta-Bleuler. It has in fact been proved that the use of an
indefinite metric is an unavoidable feature if one insists on the
preserving of causality (locality) and covariance in gauge quantum
field theories \cite{st}. This means that  one cannot restrict the
state space of the massless vector field to a Hilbert space,
the emergence of states with negative or null norm necessitates
indefinite metric quantization.

\begin{figure}[h]
\begin{center}
\input{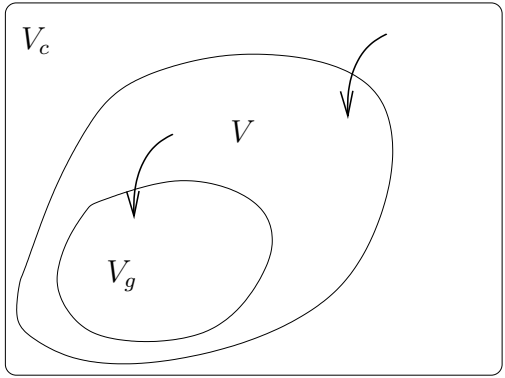}
\caption{Gupta-Bleuler structure lying behind indecomposable representation of Poincar\'e or dS group}
\label{ress1}
\end{center}
\end{figure}

An indecomposable group representation structure is needed (exactly
like for the electromagnetic field in Minkowski space) where the
physical states belong to a subspace (characterized by the
divergencelessness condition of the field operator \cite{wiga}) $V$
of solutions, but where the field operator must be defined on a
larger gauge dependent space $V_c$ (which contains negative norm
states as well), as shown in Fig.\ref{ress1}. The physical subspace
$V$ is invariant but not invariantly complemented in $V_c$. The same
feature repeats in $V$ where one finds the invariant (but again not
invariantly complemented) subspace of gauge solutions $V_g$. The
latter reveal to be orthogonal to all the elements of $V$ including
themselves \cite{gahamu}. Consequently one must eliminate them from
the physical states, by considering the physical state space as the
coset $V/V_g$. We will see that the physical states propagate on the
light-cone and correspond to vector massless Poincar\'e field in the
null curvature limit.

In previous studies,  the  massless vector field was
considered in flat coordinate system  covering only the one-half
of the dS hyperboloid \cite{bodu}. In Ref. \cite{allen1}, Allen and Jacobson
calculated the massless vector two-point functions in terms of
the geodesic distance. These functions are independent of the
choice of the coordinate system. The Hilbert space structure,
the vector field operator and the corresponding two-point function
are studied in the present paper in terms of coordinate-independent
de Sitter waves. We will adopt a very convenient value
for the gauge fixing parameter $c$ and it is not the usual Feynman
value $c=0$ \cite{ga}.  This  choice will eliminate from our solutions
additional logarithmic divergent terms which on the contrary
appear in Ref. \cite{allen1}.

The gauge invariant de Sitter vector field equation is presented
in Section $2$ in terms of the Casimir operator. We start from its
expression given in intrinsic coordinates and rewrite it by using
the ambient space formalism more convenient when  group
theoretical considerations are involved.  The Gupta-Bleuler
triplet is discussed in Section $3$. The invariant space is
defined with an indecomposable representation of the dS group.
Physical states correspond to the UIR's $\Pi^\pm_{1,1}$. It is the
central part of the indecomposable representation.

Section $4$ is devoted to the solutions of the field equation
(which we shall call de Sitter waves) in terms of a scalar field
$\phi$ and a generalized polarization vector  ${\cal E}_\alpha$,
according to the following expression
$$ {\cal K}_\alpha (x)= {\cal E}_\alpha(x,\partial)\phi (x).$$
The dS vector waves  are only locally defined since they are
singular on lower dimensional subsets  in dS space-time (they are
also multivalued in general, but not in the present case). For a
global definition, they must be viewed as distributions which are
boundary values of analytic continuations of the solutions to
tubular domains in the complexified de Sitter space  \cite{brmo}.

In Section $5$, we give two different methods for getting a
two-point function ${\cal W}_{\alpha \alpha'}(x,x')$. On one
hand  we introduce the two-point function  in terms of  vector
dS waves \cite{brmo}, on the other hand we define it  as a
maximally symmetric bivector \cite{allen1}. Of course we indicate
under which circumstances  both definition coincide.    We then
show that the two-point function satisfies the minimal conditions
of field equation, locality, covariance, normal analyticity. The
normal analyticity allows us to define the two-point function
${\cal W}_{\alpha \alpha'}(x,x')$.

Finally, in Section $6$ we construct the field operator.  We compute the commutator
and   show that the field operator  must be well chosen in order to yield a causal field.

\setcounter{equation}{0}
\section{de Sitter field equation}
\subsection{De Sitter  ambient space description}
The  de Sitter solution to the cosmological Einstein field
equation (with positive constant curvature) can be viewed as a
one-sheeted hyperboloid embedded in a five dimensionnal Minkowski
space $M^{5}$ :
\begin{equation}
X_H=\{\,x \in \setR^5 ;\,x^2=\eta_{\alpha\beta} x^\alpha x^\beta
=-H^{-2}=-\frac{3}{\Lambda}\},\;\;\alpha,\beta=0,1,2,3,4,
\end{equation}
where $\eta_{\alpha\beta}=$diag$(1,-1,-1,-1,-1)$ and $\Lambda$ is
the cosmological constant (in units $c = 1$). The de Sitter metric
is given by
\begin{equation}
ds^2=\eta_{\alpha\beta}dx^{\alpha}dx^{\beta}\mid_{x^2=-H^{-2}}=
g_{\mu\nu}^{}dX^{\mu}dX^{\nu},\;\; \nu,\mu=0,1,2,3,
\end{equation}
where $X^\mu$ are the four local space time coordinates on the dS
hyperboloid. This way of describing the dS space as a
(pseudo-)sphere in a higher-dimensional  Minkowski space
constitutes  the ambient space approach. Two crucial advantages
favor  the ambient space formalism: the expressions throughout
the present paper will have the simplest possible minkowskian-like
form (for obvious reasons),  and the link with group theory is
easily readable in this context.

In ambient space notations, a vector field ${\cal K}_{\alpha}(x)$
can  be viewed  as a homogeneous function in the
$\setR^5$-variables $x^{\alpha}$ with some arbitrarily chosen
degree $\sigma$ which therefore  satisfies:
\begin{equation}
x^{\alpha}\frac{\partial }{\partial x^{\alpha}}{\cal
K}(x)=x\cdot\partial {\cal K} (x)=\sigma {\cal K}(x).
\end{equation}
The choice for $\sigma$ will be  dictated by simplicity reasons
when one has to deal with  field equations. In the following we
set $\sigma=0$ so that the d'Alembertian operator
$\Box_{H}\equiv\nabla_{\mu}\nabla^{\mu}$ on dS space
($\nabla_{\mu}$ being the covariant derivative) coincides with the
d'Alembertian operator $\square_{\scriptscriptstyle
5}\equiv\partial^2$ on $\setR^5$. We will prove this shortly.

Of course, not every homogeneous vector field of $\setR^5$
represents a physical dS entity! In order to ensure that ${\cal
K}_{\alpha}(x)$  lies in the de Sitter tangent space-time it  also
must satisfy the transversality condition
\begin{equation}
x\cdot {\cal K}(x)=0.
\end{equation}
Given the importance of this tranversality property for dS fields
let us  introduce the symmetric,  transverse projector
$\theta_{\alpha \beta}= \eta_{\alpha \beta}+H^2x_\alpha x_\beta$
which satisfies $\theta_{\alpha\beta}\;
x^{\alpha}=\theta_{\alpha\beta}\; x^{\beta}=0$. It is the
transverse form of the dS metric in ambient space notation and it
is used in the construction of  transverse entities like  the
transverse derivative $\bar
\partial_\alpha=\theta_{\alpha \beta}\partial^\beta=
\partial_\alpha+H^2x_\alpha x. \partial$.

Since in most of  the works devoted to dS field theory the tensor
fields are written using  local coordinates, it is very important
to provide the link between the two approaches. The ``intrinsic''
vector field $A_{\mu}(X)$ is locally determined by the field
${\cal K}_{\alpha}(x)$ through the relation
\begin{equation}
\label{passage} A_{\mu}(X)=\frac{\partial x^{\alpha}}{\partial
X^{\mu}}{\cal K}_{\alpha}(x(X))\qquad  \quad {\cal K}_{\alpha}(x)=\frac{\partial X^{\mu}}{\partial
x^{\alpha}}{ A}_{\mu}(X(x)).
\end{equation}
In the same way one can show that the transverse projector
$\theta$ is the only symmetric and transverse tensor which is
linked to the dS metric $g_{\mu\nu}$  :
$$\frac{\partial x^{\alpha}}{\partial
X^{\mu}}\frac{\partial x^{\beta}} {\partial
X^{\nu}}\,\theta_{\alpha\beta}=g_{\mu\nu}\,.$$ The next step is to
explain how the covariant derivatives $\nabla$ are related to the
transverse derivative denoted by   $\bar
\partial$. In general, covariant derivatives
acting on a l-rank tensor are transformed according to
\begin{equation}
\nabla_{\mu}\nabla_{\nu}..\nabla_{\rho}h_{\lambda_{1}..\lambda_{l}}=
\frac{\partial x^\alpha}{\partial X^\mu} \frac{\partial x^\beta}
{\partial X^\nu}..\frac{\partial x^\gamma}{\partial X^\rho}
\frac{\partial x^{\eta_{1}}}{\partial X^{\lambda_{1}}}
..\frac{\partial x^{\eta_{l}}} {\partial X^{\lambda_{l}}}
\mbox{Trpr}\bar{\partial}_{\alpha}\mbox{Trpr}\bar{\partial}_{\beta}
..\mbox{Trpr}\bar{\partial}_{\gamma}{\cal
K}_{\eta_{1}..\eta_{l}}\,,
\end{equation}
where the transverse projection  defined by
\begin{equation} \left(\mbox{Trpr}
\,{\cal
K}\right)_{\lambda_{1}..\lambda_{l}}\equiv\theta^{\eta_{1}}_{\lambda_{1}}
..\theta^{\eta_{l}}_{\lambda_{l}}{\cal
K}_{\eta_{1}..\eta_{l}}\,,\nonumber
\end{equation}
guarantees  the transversality  in each index. Let us indicate how
this works for the scalar and vector fields respectively, since
these cases only will be considered in the next.

\paragraph{The scalar case}
\begin{equation*}
\nabla_{\mu}\phi= \frac{\partial x^\alpha}{\partial
X^\mu}\,\bar{\partial}_{\alpha}\phi
\quad\mbox{and}\quad\nabla_{\mu}\nabla_{\nu}\phi(X)= \frac{\partial
x^\alpha}{\partial X^\mu}  \frac{\partial x^\beta}{\partial
X^\nu}\left(
\bar{\partial}_{\alpha}\bar{\partial}_{\beta}\phi-H^{2}x_{\beta}\bar{\partial}_{\alpha}\phi\right)\,.
\end{equation*}
The d'Alembertian can be calculated
\begin{eqnarray*}
\Box_{H}\phi=g^{\mu\nu}\nabla_{\mu}\nabla_{\nu}\phi
=g^{\mu\nu}\frac{\partial
x^\alpha}{\partial X^\mu}  \frac{\partial x^\beta}{\partial
X^\nu}\left(\bar{\partial}_{\alpha}\bar{\partial}_{\beta}\phi-H^{2}
x_{\beta}\bar{\partial}_{\alpha}\phi\right)
=\theta^{\alpha\beta}
\left(\bar{\partial}_{\alpha}\bar{\partial}_{\beta}\phi-H^{2}
x_{\beta}\bar{\partial}_{\alpha}\phi\right)=\bar\partial^2\phi\,.
\end{eqnarray*}
Note that for a homogeneous function $\phi$ of degree $\sigma$ one
gets
$$\Box_{H}\phi\equiv\nabla_{\mu}\nabla^{\mu}\phi=\bar\partial^2\phi=
\partial^2\phi+3H^2\,(x\cdot\partial)\phi+H^2(x\cdot\partial)
(x\cdot\partial)\phi=\left(\Box_{\scriptscriptstyle
5}+H^2\sigma(\sigma+3)\right)\phi\,,$$ which motivates our choice
$\sigma=0$.

\paragraph{The vector case}

For a transverse vector field one easily obtains
\begin{equation}
\nabla_{\mu}A_{\nu}=\frac{\partial x^\alpha}{\partial X^\mu}
\frac{\partial x^\beta} {\partial X^\nu}
\left(\bar\partial_{\alpha} {\cal K}_{\beta}-H^2 x_{\beta}{\cal
K}_{\alpha}\right)\quad\mbox{which implies}\quad\nabla\cdot
A=\bar\partial\cdot{\cal K}.
\end{equation}
Moreover one gets
\begin{eqnarray}\label{eqn:cova}
\nabla_{\mu}\nabla_{\nu}A_{\rho}&=& \frac{\partial
x^\alpha}{\partial X^\mu} \frac{\partial x^\beta} {\partial
X^\nu}\frac{\partial x^\gamma}{\partial X^\rho}
\mbox{Trpr}\bar{\partial}_{\alpha}\mbox{Trpr}
\bar{\partial}_{\beta}{\cal K}_{\gamma} \nonumber=\frac{\partial
x^\alpha}{\partial X^\mu} \frac{\partial x^\beta} {\partial
X^\nu}\frac{\partial x^\gamma}{\partial X^\rho}
(\bar{\partial}_{\alpha}\bar{\partial}_{\beta}{\cal
K}_{\gamma}\\
&-&H^{2} \theta_{\alpha\gamma}{\cal K}_{\beta}-H^{2} x_{\beta}
\bar\partial_{\alpha}{\cal K}_{\gamma} +H^2x_{\gamma}\,{\cal
S}\left[H^2 x_\alpha{\cal K}_\beta-\bar\partial_\alpha {\cal
K}_\beta\right]),
\end{eqnarray} with ${\cal S}$ the non-normalized symmetrization operator. The
d'Alembertian becomes:
\begin{equation}\label{eq:box}
\Box_{H}
A_{\mu}=\nabla^{\lambda}\nabla_{\lambda}A_{\mu}=\frac{\partial
x^\alpha}{\partial X^\mu} \left[\bar\partial^2{\cal
K}_{\alpha}-H^2{\cal
K}_{\alpha}-2H^2x_{\alpha}\bar\partial\cdot{\cal K}\right] \;.
\end{equation}
In the following  we will recall  the ``massless''  vector field
equation on dS background and show how the ambient space formalism
is so well adapted  to the group theoretical content.

\subsection{Field equation}

The action for free ``massless'' vector fields
$A_\mu(X)$ propagating on de Sitter space reads $(\hbar=1)$
\cite{allen1}
\begin{equation}\label{eq:action}
S(A)=\int_{X_H} \frac{1}{4}\,F^{\mu\nu}F_{\mu\nu}\,d\sigma,
\end{equation}
where $F^{\mu\nu}=\nabla^\mu A^\nu-\nabla^\nu A^\mu$ and $d\sigma$
is the $O(1,4)$-invariant measure on $X_H$. The variational
principle applied to (\ref{eq:action}) yields the field equation
\begin{equation}
\nabla_\mu F^{\mu\nu}=\nabla_\mu (\nabla^\mu A^\nu-\nabla^\nu
A^\mu)=0\,.
\end{equation}
Since $\left[\nabla_{\mu},\nabla_{\nu}\right]A_{\lambda}=-H^2
\left(g_{\mu\lambda}A_{\nu}-g_{\nu\lambda}A_{\mu}\right)$ one
obtains the wave equation
\begin{equation}\label{eq:waveee}
(\Box_H +3H^2) A_\mu(X)-\nabla_\mu \nabla \cdot A(X)=0\,.
\end{equation}
This field equation is identically satisfied
by the gauge vector fields of the form $A_\mu=\nabla_{\mu}\phi$ because
of the property $\left[\Box_H\,\nabla_{\mu}-\nabla_{\mu}\Box_H\right]\phi=-3H^2
\nabla_{\mu}\phi$. Thus (\ref{eq:waveee}) is invariant under the
gauge transformation
\begin{equation}
A_{\mu} \longrightarrow
A_{\mu}'=A_{\mu}+\nabla_{\mu}\phi,
\end{equation}
where $\phi$ is an arbitrary scalar field. The wave equation with gauge fixing parameter $c$  reads
\begin{equation}\label{eq:wavee}
(\Box_H +3H^2) A_\mu(X)-c\nabla_\mu \nabla \cdot A(X)=0\,.
\end{equation}
Our aim is now to write the field equation (\ref{eq:wavee})  in
terms of the Casimir operator of the dS group SO$_0(1,4)$.

\subsection{Casimir operators in the field equation}

The kinematical group of the de Sitter space is the $10$-parameter
group SO$_0(1,4)$ (connected component of the identity in O$(1,4)$), which is one of the two possible deformations of the Poincar\'e
group. There are two Casimir operators
\begin{equation}
Q^{(1)}_1=-\frac{1}{2}L_{\alpha\beta}L^{\alpha\beta},\qquad
Q^{(2)}_1=-W_{\alpha}W^{\alpha},\label{eq:cas}
\end{equation}
where
\begin{equation}
W_{\alpha}=-\frac{1}{8}\epsilon_
{\alpha\beta\gamma\delta\eta}L^{\beta\gamma}L^{\delta\eta},
\quad\mbox{with  10 infinitesimal generators}\quad
L_{\alpha\beta}=M_{\alpha\beta}+S_{\alpha\beta}.
\end{equation}
The subscript $1$ in $Q^{(1)}_1$, $Q^{(2)}_1$ reminds that the
carrier space is constituted by  vectors. The orbital part
$M_{\alpha\beta}$, and the action of the spinorial part
$S_{\alpha\beta}$ on a vector field ${\cal K}$ defined on the
ambient space read respectively \cite{ta}
\begin{equation}
M_{\alpha\beta}=-i (x_\alpha\partial_\beta-x_\beta\partial_\alpha),\qquad
S_{\alpha\beta}{\cal K}_{\gamma}=-i(\eta_{\alpha\gamma}{\cal
K}_{\beta}-\eta_{\beta\gamma} {\cal
K}_{\alpha}).
\label{eq:spi}
\end{equation}
The symbol $\epsilon_{\alpha\beta\gamma\delta\eta}$ holds for the
usual antisymmetrical tensor. The action of the Casimir operator
$Q_1^{(1)}$  on ${\cal K}$ can be written in the more explicit
form
\begin{equation}\label{eq:act}
Q_1^{(1)}{\cal K}(x)=\left(Q_{0}^{(1)}-2\right){\cal K}(x)+2
x\;\bar\partial\cdot{\cal K}(x)-2  \partial\; x\cdot{\cal K}(x),
\end{equation}
where, $Q_{0}^{(1)}=-{{1}\over{2}}M_{\alpha\beta}M^{\alpha\beta}$
is the scalar Casimir operator. We are now in position to express
the wave equation (\ref{eq:wavee}) by using the Casimir operators.
This can be done with the help of equation (\ref{eq:box}) since
$Q_{0}^{(1)}=-H^{-2}(\bar\partial)^2$. The d'Alembertian operator
becomes
\begin{equation}
\Box_{H}
A_{\mu}=\nabla^{\lambda}\nabla_{\lambda}A_{\mu}=-\frac{\partial
x^\alpha}{\partial X^\mu} \left[Q_{0}^{(1)}H^2{\cal
K}_{\alpha}+H^2{\cal
K}_{\alpha}+2H^2x_{\alpha}\bar\partial\cdot{\cal K}\right] \;,
\end{equation}
and the  equation (\ref{eq:wavee}) with this new notation reads
\begin{equation}
\left(Q_0^{(1)}-2\right){\cal K}(x)+2x\,\bar
\partial\cdot {\cal K}(x)+cH^{-2}\bar \partial
\,\partial\cdot {\cal K}(x)=0\,.
\end{equation}
Finally  using (\ref{eq:act}) one obtains the field equation formulated in terms of the Casimir operator $Q_1^{(1)}$ :
\begin{equation}\label{eq:wave}
Q_1^{(1)}{\cal K}(x)+cD_1
\partial\cdot {\cal K}(x)=0,\qquad\mbox{where}\quad D_1=H^{-2}\bar
\partial\,.
\end{equation}
But, as we will see, the ``minimal" (or optimal) choice of $c$ is not zero,
contrary to the flat space case (Feynman  gauge).
This is because the choice $c=0$  yields logarithmic divergent
terms in the vector field expression. The ``minimal" choice on the
contrary is chosen so that it allows to eliminate those terms. Before coming back
to this point, let us turn to the group-theoretical content of
this equation.

\subsection{Group theoretical notions}

The operator $Q_1^{(1)}$ which commutes with the action of the group
generators can be used to classify the UIR's {\it i.e.,}
\begin{equation}\label{rep}
(Q_1^{(1)}-\langle Q_1^{(1)}\rangle){\cal K}(x)=0.
\end{equation}
Following Dixmier in Reference \cite{dix} we get a classification
scheme by using a pair $(p,q)$ of parameters involved in the
following possible spectral values of the Casimir operators:
\begin{equation}
Q^{(1)}=\left(-p(p+1)-(q+1)(q-2)\right)I_d ,\qquad\quad
Q^{(2)}=\left(-p(p+1)q(q-1)\right)I_d\,.
\end{equation}
As comprehensively described in Appendix \ref{appA}, three types of
scalar, tensorial or spinorial UIR are distinguished for
SO$_{0}(1,4)$ according to the range of values of the parameters $q$
and $p$ \cite{dix,tak}, namely the principal, the complementary and
the discrete series. In the following, we shall restrict  the list
to those among all unitary representations which precisely have a
minkowskian physical spin-$1$ interpretation in the limit $H=0$. The
flat limit tells us that for the principal and the complementary
series it is the value of $p$ which has a spin meaning, and that, in
the case of the discrete series, the only representations which have
a physically meaningful minkowskian counterpart are those with $p=q$
. The spin-$1$ tensor representations relevant to the present work
are the following :
\begin{itemize}
\item[i)] The UIR's $U^{1,\nu}$ in the principal series where
$p=s=1$ and $q=\frac{1}{2 }+i\nu$ corresponds to the Casimir
spectral values:
\begin{equation}
\langle Q^{(1)}_1\rangle=\nu^2+\frac{1}{4},
\end{equation}
with the parameter $\nu \in \setR$ (note that $U^{1,\nu}$ and
$U^{1,-\nu}$ are equivalent). The principal series corresponds to
the massive case \cite{gata}.
\item[ ii)] The UIR's $V^{1,q}$ in the complementary series where
$p=s=1$ and $q=\frac{1}{2}+\nu,$ corresponds to
\begin{equation}
\langle
Q_1^{(1)}\rangle=\frac{1}{4}-\nu^2,\qquad\mbox{with}\qquad\;0<\vert\nu\vert<\frac{1}{2},\quad\mbox{and}\quad
\nu \in \setR.
\end{equation}

\item[iii)] The UIR's $\Pi^{\pm}_{1,1}$ in the discrete series where
$q=p=s=1$ correspond to
\begin{equation}
\langle Q_1^{(1)}\rangle=0 .
\end{equation}
\end{itemize}
By comparing  equations  ($\ref{eq:wave}$) and  ($\ref{rep}$), it is
immediately seen that the spin-$1$  massless field in de Sitter
space corresponds to the elements  $\Pi^{\pm}_{1,1}$ of the discrete
series with the Casimir operator  eigenvalue $\langle
Q_1^{(1)}\rangle=0$. It is shown in \cite{barutbohm} that there are
exactly two inequivalent UIR's of the de Sitter group SO$_0(1,4)$
which extend biunivocally to the conformal group SO$_0(2,4)$, namely
$\Pi^{\pm}_{1,1}$.   These two unitary irreducible representations
differ in the sign of a helicity-like eigenvalue related to the
representation of the subgroup SO$(3)$ which is left unchanged after
zero-curvature limit, \emph{i.e.} the subgroup of space isotropy. It
is therefore reasonable to say that both representations are
distinguished according to their helicity represented by  the symbol
$\pm$.

These representations are associated to the subspace of solutions to
Eq. (\ref{eq:wave}) characterized by $\partial\cdot {\cal K}=0$.
Thus, it is natural to use the solution of the equation
\begin{equation}
(Q_1^{(1)}-\langle Q_1^{(1)}\rangle){\cal K}(x)=0\,,
\end{equation}
already given in \cite{gata} for the massive case. The
corresponding vector field  solution can be put under the form
\begin{equation}
{\cal K}_{\alpha}(x)={\cal E}_{1\alpha }(x,\xi)\phi
(x)+\frac{1}{\langle Q_1^{(1)}\rangle} \;{\cal E}_{2\alpha
}(x,\xi)\phi (x)\,,
\end{equation}
where ${\cal E}_{1\alpha }(x,\xi)$, ${\cal E}_{2\alpha }(x,\xi)$
and $\phi (x)$  also contain constant terms involving the
parameters $p$ and $q$, but which do not diverge for the specific
values $p=q=1$ corresponding to  the massless vector UIR ($\langle
Q_1^{(1)}\rangle=0$). Clearly, a singularity appears for the spin
$1$ massless field due to the term $1/\langle Q_1^{(1)}\rangle$.
The subspace determined by $\partial\cdot {\cal K}=0$ considered
so far is therefore not sufficient for the construction of a
quantum massless vector field. One must solve the equation in a
larger space which includes the $\partial\cdot{\cal K}\neq 0$
types of solutions. As expected, one finds three main types of
solutions:  the divergencelessness type, the gauge type and the
latter solutions which aren't divergenceless.

\setcounter{equation}{0}
\section{The Gupta-Bleuler triplet}\label{GBT}
As stated in \cite{binegar}, ``the appearance of [the
Gupta-Bleuler] triplet seems to be universal in gauge theories,
and crucial for quantization". The ambient space formalism will
allow to exhibit this triplet for the present field in exactly the
same manner as it occurs for the electromagnetic field. The
Gupta-Bleuler structure of the latter is reminded in Appendix \ref{appB}.

We start with  the field equation $(\ref{eq:wave})$. The following
dS invariant bilinear form (or inner product) on the space of
solutions is defined for two modes (which we also note $\K_1$, $\K_2$
whatever their depending on a specific dS coordinate system) in \cite{gahamu} as
\begin{equation} \label{produitsca}
(\K_1,\K_2)=\frac{i}{H^2}\int_{\renewcommand{\arraystretch}{0.4}
\!\!\begin{array}{l}
\begin{scriptstyle}S^3\end{scriptstyle}\\
\begin{scriptstyle}\rho=0
\end{scriptstyle}\end{array}}\!\!\! \left[\K_{1}^*\cdot
{\partial}_{\rho} \K_2-c(({\partial}_{\rho} x)\cdot
\K_{1}^*)(\partial\cdot \K_2)-(1^*\rightleftharpoons
2)\right]d\Omega,\end{equation}
where we have used the  system of bounded global intrinsic
coordinates $(X^\mu,\;\mu=0,1,2,3)$ well-suited to describe a
compactified version of dS space, namely S$^3 \times{\rm S}^1$.
Let us recall that this coordinate system, known as conformal
coordinates, is defined by
\begin{equation}\label{coordinates}
\left\{\begin{array}{rcl} x^0&=&H^{-1}\tan\rho\\
x^1&=&(H\cos\rho)^{-1}\,(\sin\alpha\,\sin\theta\,\cos\varphi),\\
x^2&=&(H\cos\rho)^{-1}\,(\sin\alpha\,\sin\theta\,\sin\varphi),\\
x^3&=&(H\cos\rho)^{-1}\,(\sin\alpha\,\cos\theta),\\
x^4&=&(H\cos\rho)^{-1}\,(\cos\alpha),\end{array}\right.
\end{equation}
where $-\pi/2<\rho<\pi/2$, $0\leq\alpha\leq\pi$,
$0\leq\theta\leq\pi$ and $0\leq\varphi < 2\pi$. For the fields
that satisfy the divergencelessness condition, the inner product
becomes c independent and KG-like:
\renewcommand{\arraystretch}{0.6}
$$(\K_1,\K_2)=\frac{i}{H^2}\int_{\renewcommand{\arraystretch}{0.4}
\!\!\begin{array}{l}
\begin{scriptstyle}S^3\end{scriptstyle}\\
\begin{scriptstyle}\rho=0
\end{scriptstyle}\end{array}}\!\!\!
[\K_{1}^*\cdot{\partial}_{\rho} \K_2-\K_{2}\cdot{\partial}_{\rho}
\K_{1}^*]d\Omega.$$
\renewcommand{\arraystretch}{1}

Let us now define the Gupta-Bleuler triplet $V_g \subset V \subset
V_{c}$ carrying the undecomposable structure for the  unitary
representation of the de Sitter group appearing in our problem.

\begin{itemize}

\item[-] The space  $V_{c}$ is the space  of all square integrable
(with respect to (\ref{produitsca})) solutions of the field
equation (\ref{eq:wave}), including negative norm solutions. It is
c dependent so that one can actually adopt an optimal value of c
which eliminates logarithmic divergent solutions \cite{gahamu}. In
the next section, we will show  that this  value is
$c=\frac{2}{3}$, (more generally for a spin s field,
$c=(2/(2s+1))$ \cite{ga}).

\item[-] It contains a closed subspace $V$ of solutions
satisfying the divergencelessness condition. This invariant subspace
$V$ is not invariantly complemented in $V_c$. In view of Eq.
(\ref{eq:wave}), it  is obviously c independent.

\item[-]The subspace $V_g$ of $V$ consists of the gauge  solutions of the form
${\cal K}_g=D_1\phi_M$. These are orthogonal to every element in
$ V$ including themselves. They form an invariant subspace of $V$
but admit no  invariant complement in $V$.
\end{itemize}

The inner product is indefinite in $V_{c}$, semi-definite in $V$
and is positive definite in the quotient space $V/V_{g}$. The
latter is the physical state space. The de Sitter group acts on
the physical  (or transverse) space $V/V_{g}$ through the
massless, helicity $\pm 1$ unitary representation $ \Pi^+_{1,1}
\bigoplus \Pi^-_{1,1}$. We now   characterize the gauge state
space $V_{g}$ and the scalar states belonging to  the space
$V_{c}/V$. \vspace{0.1cm}

\subsection{The gauge states:}

With a solution of the form ${\K}_g=D_1\phi_M$, equation
(\ref{eq:wave}) becomes (using $D_1Q_0\phi=Q_1D_1\phi$)
\begin{equation}
(1-c)D_1Q^{}_0 \phi_M=0\,.
\end{equation}
At this stage one must distinguish the two cases $c=1$ and $c\neq
1 $.

\begin{itemize}
\item If $c=1$, the scalar field  $\phi_M$ is unrestricted, let alone mild differentiability conditions, and the
gauge state space is given by vectors of the form
$D_1\phi\;$ for a differentiable scalar field $\phi$.

\item If $c\neq 1$, it is seen that  $\phi_{M}$ corresponds to a
massless minimally coupled scalar field characterized by
$Q_0\phi_M=0$ (possibly up to the addition of a particular
solution of the inhomogeneous equation $Q_0\phi_M= \mathrm{cst}$, and
associated with the representation $\Pi_{1,0}$).  Moreover, since
one has $L_{\alpha\beta}D_1\phi_M=D_1 M_{\alpha\beta}\phi_M$, this
shows that the  vector $K_{g}$ does not carry any spin. Thus  it
is entirely characterized by its scalar content and can be
associated to $\Pi_{1,0}$. Note that the representation structure
of the minimally coupled scalar field requires another
Gupta-Bleuler type of triplet where the gauge states are  the
constant fields \cite{gareta1}.
\end{itemize}

\subsection{The scalar states:}

The scalar states belong to the quotient space $V_{c}/V$. In order
to characterize them,  let us take the divergence of Eq.
(\ref{eq:wave}):
\begin{equation}\label{eqdiv}
0=\bar\partial\cdot\left(Q_1\K(x)+cD_1
\partial\cdot
\K(x)\right)=Q_0^{}\bar\partial\cdot \K
(x)+cH^{-2}\bar\partial^{2}\bar\partial\cdot \K(x)\;,
\end{equation}
from which one derives
\begin{equation}
\label{eqdiv}
(1-c)Q^{}_0 \bar\partial\cdot \K(x)=0\,.
\end{equation}
Again one must distinguish between $c=1$ and $c\neq 1$.

\begin{itemize}

\item If $c =1$, the vector $\K(x)$ is  unrestricted except
obvious differentiability conditions. For this special value of $c$
one loses the opportunity of restraining the  space $V_c$.

\item
If $c\neq 1$, the divergence $\bar\partial\cdot\K(x)$ again
correspond to a massless minimally coupled scalar field associated
with the representation $\Pi_{1,0}$.
\end{itemize}

\begin{figure}[h]
\begin{center}
\input{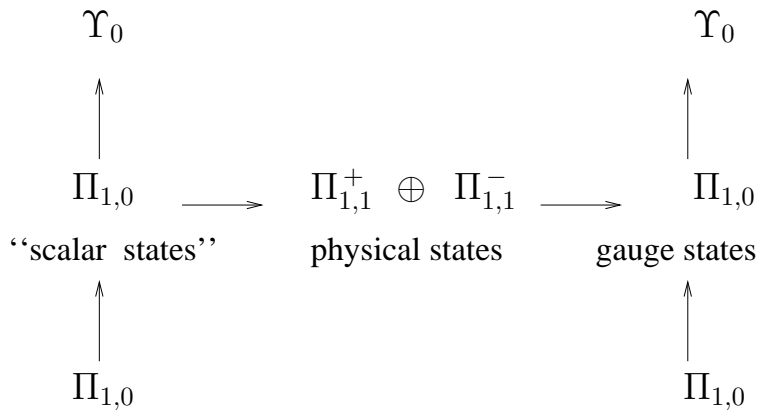}
\caption{Massless vector field indecomposable group representation structure.}\label{mn1}
\end{center}
\end{figure}

The representation structure of the full space $V_c$, can be
pictured as shown in Figure \ref{mn1}.

The arrows indicate the leaks under the de Sitter group action.
The central parts $\Pi^\pm_{1,1}$ are the only spin one unitary
irreducible representations of the dS group that admit a
minkowskian massless spin 1 interpretation (due to their conformal
invariance). A closer look at the scalar and the gauge states
solutions  reveals a further Gupta-Bleuler triplet described in
details in  Ref. \cite{gareta1}. Indeed, the scalar and gauge
states are associated to the representation $\Pi_{1,0}$ (minimally
coupled scalar field) where the space of constant functions
assumes the role of gauge space which carries the trivial UIR
$\Upsilon_0$ (on which both Casimir operator vanish). The coset
spaces (scalar states)/(constant functions) and (gauge
states)/(constant functions) are  spaces which contain negative
norm states and  which  also carry the UIR $\Pi_{1,0}$.

In the following we present the solutions of Eq. (\ref{eq:wave})
and explicitly compute the group actions indicated in the above discussion and illustrated by the figure \ref{mn1}.

\setcounter{equation}{0}
\section{de Sitter vector waves, field equation solution}

We now solve the ``massless'' vector wave equation with gauge
fixing term. This equation   reads
\begin{equation}\label{eq:wave1}
\left(Q_1-\langle Q_1\rangle\right)\K(x)+cD_1 \bar\partial\cdot
\K(x)=0\qquad\mbox{with} \qquad \langle Q_1\rangle=0.
\end{equation}
A general solution can be written in terms of two scalar fields
\begin{equation}\label{eq:ansatz}
\K=\bar Z\phi+D_{1}\phi_1,
\end{equation}
where $Z$ is a constant vector  and $\bar
Z_{\alpha}=\theta_{\alpha\beta} Z^{\beta}$. The scalar field
$\phi_1$ is defined up to the addition of a scalar field $\phi_g$.
After inserting $\K(x)$ in (\ref{eq:wave1}) and with the help of
the following relations
\begin{equation}\label{eq:relation1}
Q_1D_1\phi=D_1Q_0\phi ,\;,\qquad Q_1\bar Z_{\alpha}\phi=\bar
Z_{\alpha}(Q_0-2)\phi-2H^2{D_1}_{\alpha}(x\cdot Z)\phi,
\end{equation}
one finds from the linear independence of the terms in
(\ref{eq:ansatz}) that
\begin{align}\label{systeme}
&\left( Q_0-2\right)\phi=0, \\
\label{systeme1}&  Q_{0}^{}\phi_1-2H^2(x\cdot Z) \phi+
c\;\bar\partial\cdot\K =0 \,.
\end{align}
Equation (\ref{systeme}) means that the scalar field $\phi$  obeys
\begin{equation}\label{eq:scalar}
(\Box_H+2H^2)\phi=0\,.
\end{equation}
It therefore corresponds to the  ``massless'' conformally coupled scalar
field \cite{brmo,ta}. This equation is invariant under conformal transformations,
and its  solutions  are known to be the dS ``massless'' waves
\begin{equation}
\phi(x)=(Hx\cdot\xi)^{\sigma}\quad\mbox{with}\quad \sigma=-1,-2.
\end{equation}
These are defined on connected open subsets of $X_{H}$ such that
$x \cdot \xi \neq 0$, where $\xi \in \setR^5 $ lies on the null
cone ${\cal C} = \{ \xi \in \setR^5;\;\; \xi^2=0\}$. They are
homogeneous with degree $\sigma$ on ${\cal C}$ and thus are
entirely determined by specifying their values on a well chosen
curve (the orbital basis ) $\gamma$ of ${\cal C}$. As such, the dS
scalar waves are not square integrable. However, physical de
Sitter entities like square integrable states can be built as
superpositions of such waves (by making $\xi$ vary in ${\cal C}$). They play in de Sitter space, the
role of the plane waves in Minkowski space.

Now since the
divergence of (\ref{eq:ansatz}) reads
\begin{equation}\label{div}
\bar\partial \cdot\K=-Q_0 \phi_1 +Z\cdot \bar \partial
\phi+4H^2x\cdot Z \phi\,, \end{equation} one obtains from equation
(\ref{systeme1}) that the  scalar field $\phi_1$ satisfies
\begin{equation}\label{eq:q0phig}
Q_0\phi_1 =-\frac{c}{1-c}\left(H^2x\cdot Z \phi+Z\cdot\bar
\partial \phi\right)+\frac{2-3c}{1-c}\, Z\cdot \bar
\partial \phi\,.
\end{equation}
At this stage we could fix the value of c. However it is
interesting to consider the general case in order to see in which
way a specific value of c correspond to the simplest case.

\subsection{\bf The general case}
Our task is to invert the equation (\ref{eq:q0phig}) in order   to
completely determine $\phi_1$ in terms of the conformally coupled
scalar field $\phi$. According to Eq. (\ref{eq:q0phig}), $\phi_1$
will be entirely  determined by $\phi$, except for an additional
term $\phi_g=Q_0^{-1}(0)$. First of all, one can put the equation
(\ref{eq:q0phig}) in the form
\begin{eqnarray}
\phi_1 &=&Q_0^{-1}\left(-\frac{c}{1-c}(H^2x\cdot Z
\phi+Z\cdot\bar\partial \phi)+\frac{2-3c}{1-c} H^2x\cdot Z \phi
\right)+Q_0^{-1}(0)\,,\nonumber\\
 &=&-\frac{c}{2(1-c)}\left(H^2x\cdot Z
\phi+Z\cdot\bar\partial
\phi\right)+\frac{2-3c}{1-c}H^2Q_0^{-1}x\cdot Z
\phi+Q_0^{-1}(0)\,. \label{qophi}
\end{eqnarray}
This can be verified using the relations
\begin{eqnarray}
Q_{0}H^2 x\cdot Z \phi &=&H^{2} x\cdot Z\left(
Q_0-4\right)\phi-2Z\cdot \bar\partial\phi=-2H^2 x\cdot Z
\phi-2Z\cdot
\bar\partial\phi\,,\label{rel1}\\
Q_{0}\,Z\cdot \bar\partial\phi&=&Z\cdot \bar\partial\left(
Q_0+2\right)\phi+2H^2 x\cdot Z \,Q_0\phi=4Z\cdot \bar\partial\phi
+4H^2 x\cdot Z \phi\,,\label{rel2}
\end{eqnarray}
which imply
\begin{equation}\label{rel3}
 Q_0\left(H^2 x\cdot Z + Z\cdot \bar\partial \right)\phi=2
 \left(H^2 x\cdot Z + Z\cdot \bar\partial\right)\phi\,.
 \end{equation}
Note also that the term $Q_0^{-1}(0)$ is a minimally coupled
scalar field  $\phi_{M}$ which satisfies the equation
$Q_0\phi_M=0$. Then for the vector mode $\K(x,\xi)$  the general
solution reads
\begin{equation}\label{sol}
\K(x,\xi,Z)=\bar Z \phi-\frac{c}{2(1-c)}D_1\left(H^2x\cdot Z
\phi+Z\cdot\bar\partial
\phi\right)+\frac{2-3c}{1-c}H^2D_1Q_0^{-1}x\cdot Z
\phi+HD_1\phi_M.
\end{equation}

Contrary to the minkowskian QED, the simplest gauge fixing  is not
the Feynman type of choice $c=0$, which here would yield
\begin{equation}\label{eq:gauge}
\K(x,\xi,Z)=\bar Z \phi+2HD_1Q_0^{-1}Hx\cdot Z \phi+HD_1\phi_M\,.
\end{equation}
Actually, the term $Q_0^{-1}x\cdot Z \phi$ bears a singularity in
the solution \cite{berotata}
$$Q_0^{-1}x\cdot Z(x.\xi)^{\sigma}=\frac{-1}{(\sigma+1)(\sigma+4)}x\cdot
Z(x.\xi)^{\sigma}, \; \sigma=-1.$$ The third term in (\ref{sol}),
which is responsible for the appearance of a singularity, is removed
after choosing
 $c=2/3$. These solutions will correspond to what we call ``minimal
case''.

\subsection{\bf The minimal case, $c=2/3$}
For the choice  $c=\frac{2}{3}$, and according to (\ref{qophi}),
$\phi_1$ is determined in terms of $\phi$ and $\phi_M$ in the
following way
\begin{equation}
\phi_1 =-Z\cdot\bar \partial \phi-H^2 x\cdot Z \phi+H\phi_M\;.
\end{equation}
Now starting with formula (\ref{sol}), the ``massless'' vector
wave in this gauge can be  written  in terms of a generalized
polarization  vector ${\cal E}_{\alpha}(x,\xi,Z)$,   a
``massless'' conformally coupled scalar field $\phi$ and gauge
solutions as
\begin{align}\label{solution}
\nonumber \K_{\alpha}(x,\xi,Z)&=\left[\bar Z_{\alpha}- D_{1\alpha}\left(Z\cdot\bar
\partial+H^2 x\cdot Z\right)\right]\phi(x)+HD_{1\alpha}\phi_M\\
& \equiv{\cal E}_{\alpha}(x,\xi,Z)\,\phi(x)+HD_{1\alpha}\phi_M\,,
\end{align}
with the  polarization vector
\begin{equation}
{\cal E}_{\alpha}(x,\xi,Z)= \left[-\sigma\bar
Z_{\alpha}-\sigma(\sigma+1)\frac{Z\cdot x}{x\cdot\xi} \,  \bar
\xi_{\alpha}- \sigma(\sigma-1)\frac{Z\cdot\xi} {(H
x\cdot\xi)^2}\,\bar \xi_{\alpha}\right].
\end{equation}
The two possible solutions for $\K(x,\xi,Z)$ corresponding to
$\sigma=-1,-2$ are
\begin{align*}
\label{}
     \K_{1\alpha}(x,\xi,Z)=&{\cal
E}_{1\alpha}(x,Z,\xi)(Hx\cdot\xi)^{-1}+D_1\phi_M, \\
     \K_{2\alpha}(x,\xi,Z)=&{\cal
E}_{2\alpha}(x,Z,\xi)(Hx\cdot\xi)^{-2}+D_1\phi_M,
\end{align*}
 where
 \begin{align*}
\label{}
    {\cal E}_{1\alpha}(x,\xi,Z)=&\bar Z_{\alpha}- 2 \frac{Z\cdot \xi}{(H
x\cdot\xi)^2} \,\bar \xi_{\alpha},   \\
     {\cal E
}_{2\alpha}(x,\xi,Z)=&2\bar Z_{\alpha}-2\frac{Z\cdot x}{x\cdot \xi}
\,\bar \xi_{\alpha}- 6 \frac{Z\cdot \xi}{(H x\cdot\xi)^2} \bar
\xi_{\alpha}.
\end{align*}
We are now in position to rewrite the general
vector wave solution in the convenient following form
\begin{equation}
\K^c=\K^{\frac{2}{3}}+\frac{\frac{2}{3}-c}{H^2(1-c)}\bar
\partial Q_0^{-1}\bar\partial\cdot \K^{\frac{2}{3}},
\end{equation}
where the $\K^{\frac{2}{3}}$ is the field solutions for
$c=\frac{2}{3} $. This can be checked using the relations
$(\ref{div})$, $(\ref{rel1})$ and $(\ref{rel2})$. The term
$Q_0^{-1}\bar\partial\cdot \K^{\frac{2}{3}}$ is responsible for
the singularity.
In the next, we shall work essentially with the $c=2/3$ gauge.

\subsection{\bf Group action and physical subspace}

In this part we would like to make more complete the description
given in Section \ref{GBT} of the  Gupta-Bleuler structure. More
precisely, we characterize the various solutions appearing in the
Gupta-Bleuler triplet. The elements of $V_{g}$ and $V_{c}$ are
already known. These are respectively the gauge solutions
$D_{1}\phi_M$ and the vectors defined by $(\ref {sol})$. The
subspace V of solutions is characterized by the divergencelessness
condition. The divergence of $\K(x,\xi,Z)$ defined in equation
$(\ref {sol})$ is given by
\begin{eqnarray}
\bar\partial\cdot\K &=&\bar\partial\cdot\bar Z
\phi+\frac{c}{2(1-c)}\;Q_0\left(H^2x\cdot Z
\phi+Z\cdot\bar\partial \phi\right)-\frac{2-3c}{1-c}\,H^2x\cdot Z
\phi-H\underbrace{Q_0\phi_M}_{=\;0}\\
&=& \frac{1}{1-c}\left[(2+\sigma)\;H^2 Z\cdot x +\sigma
\frac{Z\cdot \xi}{x\cdot \xi}\right]\phi(x)\,,
\end{eqnarray}
where we have used the relations $\bar\partial\cdot \bar{Z}=4H^2
x\cdot Z$, and
$\bar\partial\left(Hx\cdot\xi\right)^{\sigma}=\sigma
H\bar\xi\left(Hx\cdot\xi\right)^{\sigma-1}$ with  Eq.(\ref{rel3}).
As expected, it is seen that the divergencelessness subspace of
solutions is not determined by a specific choice of c (for $c\neq
1$). We therefore can choose to work with the special value
$c=2/3$. More important, one sees that the subspace $V$ will be
characterized by the conditions
\begin{equation}
\sigma=-2\qquad\mbox{and}\qquad Z\cdot\xi=0\,.
\end{equation}
The value $\sigma=-2$ selects the family $\K_{2\alpha}(x,\xi,Z)$ and $Z\cdot\xi = 0$ is a condition on the polarization five-vector $Z$. Therefore the elements of V  will be made of superpositions of the following solutions:
$$
\K_{2\alpha}(x,\xi,Z)={\cal
E}_{2\alpha}(x,Z,\xi)(Hx\cdot\xi)^{-2}+HD_1\phi_M\qquad\mbox{with}\qquad
{\cal E }_{2\alpha}(x,\xi,Z)=2\left(\bar Z_{\alpha}-\frac{Z\cdot
x}{x\cdot \xi} \,\bar \xi_{\alpha}\right)\,.$$

We have indicated that the Gupta-Bleuler triplet is based on three invariant spaces of
solutions. Let us first show that each one of these spaces is effectively invariant.
It is trivial that the elements of $V_c$ for a given $c$ remain in
$V_c$ under the group action. Moreover, the gauge  solutions
also form a closed subspace since
$L_{\alpha\beta}D_1\phi_M=D_1M_{\alpha \beta }\phi_M $. In order
to show that $V$ is invariant under the group action, let us
consider the infinitesimal group action.  The infinitesimal group
action reads
\begin{eqnarray}
L_{\alpha\beta}^{(1)} \K_{\gamma}=L_{\alpha\beta}^{(1)}\left(\bar
Z\phi+D_{1}\phi_1 \right)=\bar Z\,M_{\alpha\beta}\phi-i \left(
Z_{\beta}\theta_{\alpha\gamma}-
Z_{\alpha}\theta_{\beta\gamma}\right)\phi+D_{1}\,M_{\alpha\beta}\phi_1\,,
\end{eqnarray}
and, with
\begin{eqnarray}
M_{\alpha\beta}^{} (Z\cdot x)\phi&=&(Z\cdot x)
M_{\alpha\beta}\phi-i \left( Z_{\beta}x_{\alpha}-
Z_{\alpha}x_{\beta}\right)\phi\,,\nonumber\\
M_{\alpha\beta}\, Z\cdot \bar\partial\phi&=&Z\cdot \bar\partial
M_{\alpha\beta}\phi\,-i \left( Z_{\beta}\bar\partial_{\alpha}-
Z_{\alpha}\bar\partial_{\beta}\right)\phi\,,
\end{eqnarray}
one obtains
\begin{eqnarray}
L_{\alpha\beta}^{(1)} \K_{\gamma}
&=&\bar{Z}_{\gamma}M_{\alpha\beta}\phi -D_1\left(Z\cdot\bar
\partial  +H^2 x\cdot Z \right) M_{\alpha\beta} \phi+HD_1M_{\alpha\beta}\phi_M
\nonumber\\
&-&i\left( x_{\beta}Z_{\alpha}-
x_{\alpha}Z_{\beta}\right)\bar\partial_{\gamma}\phi-iH^{-2}\left(
Z_{\alpha}\bar\partial_{\gamma} \bar\partial_{\beta}-
Z_{\beta}\bar\partial_{\gamma}\bar\partial_{\alpha}\right)\phi\,.
\end{eqnarray}
It is possible to make explicit  the divergence of the latter
equation
\begin{align}
\nonumber \bar\partial^{\gamma}L_{\alpha\beta}^{(1)} \K_{\gamma}
=&3iH^{2}(\sigma+2)\left(x_\beta
Z_\alpha-x_{\alpha}Z_{\beta}\right)\left(Hx\cdot\xi\right)^{\sigma}\\
-&3iH^{2}\sigma(\sigma+2)\left(x\cdot Z \right)\left(x_\beta
\xi_\alpha-x_{\alpha}\xi_{\beta}\right)\left(Hx\cdot\xi\right)^{\sigma-1}
\nonumber\\
+&3iH^{2}\sigma(\sigma -1)\left(\xi\cdot Z \right)\left(x_\beta
\xi_\alpha-x_{\alpha}\xi_{\beta}\right)\left(Hx\cdot\xi\right)^{\sigma-2}\,,
\end{align}
which  shows  that the subspace of divergencelessness
solutions with $\sigma=-2$ and $Z\cdot\xi$ is invariant. Because of
divergencelessness and  transversality   $x\cdot{\cal E
}_{2\alpha}(x,\xi,Z)=0$, one finds that the independent components
are reduced from the original five to three. The actual physical
subspace  of solutions, that is the quotient space $V/V_{g}$,
corresponds to the two-component polarization vectors ${\cal E
}_{2\alpha}^{\lambda}(x,\xi,Z)$,  $\lambda=1,2$. The latter
satisfy a kind of transversality condition
\begin{equation}\label{conc}
\bar\xi\cdot{\cal E }_{2\alpha}^{\lambda}(x,\xi,Z)=\xi\cdot{\cal E }_{2\alpha}^{\lambda}(x,\xi,Z)=0.
\end{equation}
Note that  transversality relations (\ref{conc}) are
valid only for the physical states in $V/V_g$. They no longer hold for
states belonging to $V$:  by adding  the gauge solutions
$D_1\phi_M$ with for instance
$\phi_M=\left(Hx\cdot\xi\right)^{-3}$ ($\sigma=-3$ corresponds to
a minimally coupled scalar field) one obtains
$$
\bar\xi\cdot
D_{1}\phi_M=\bar\xi\cdot\bar\xi \;H^{-1}
\left(Hx\cdot\xi\right)^{-4}= H^{-1}
\left(Hx\cdot\xi\right)^{-2}\neq 0\qquad\mbox{ since}\qquad\bar\xi\cdot\bar\xi
=\left(Hx\cdot\xi\right)^{2}.$$

\subsection{\bf General comments concerning the dS vector waves}
 An important difference with
the minkowskian case is that the polarization vectors ${\cal
E}_{i\alpha}(x,\xi,Z)\, i = 1,2$ are function of the space-time variable $x$.
Moreover, unlike the minkowskian and  the dS ``massive'' vector
cases, these two solutions, in our notations, are not complex
conjugate of each other. Note that they satisfy the homogeneity properties
\begin{equation} {\cal
E}_{\alpha}(x,a\xi,Z)={\cal
E}_{\alpha}(x,\xi,Z)\quad\mbox{and}\quad {\cal
E}_{\alpha}(ax,\xi,Z)={\cal E}_{\alpha}(x,\xi,Z), \nonumber
\end{equation}
and thus the dS waves $\K_{\alpha}(x,\xi,Z)$ are homogeneous with
degree $\sigma$ as functions of $\xi$ on the null cone ${\cal C}$ as well as on the dS
submanifold $X_{H}$. Also note that as  functions on $\setR^{5}$, these
waves are homogeneous with degree zero since in that case $H$
becomes a function of $x$: $H(x)=-1/\sqrt{- x\cdot x }$.

The arbitrariness introduced with the constant vector $Z$ will be
removed  by comparison with the minkowskian case. Unfortunately,
our notations for the ``massless'' conformally coupled scalar
waves are not adapted to the computation of the $H=0$ limit. It is
due to the fact that contrary to the ``massive case''  the values
$\sigma=-1,-2$ are constant \cite{brepmo}. In order to get a hint
of the behavior of the field equation solutions in the limit $H=0$
(at least the scalar part), one can use the conformal coordinate
system which has been introduced in Eq. (\ref{coordinates}).
The square-integrable  solutions of the field equation
(\ref{eq:scalar}) are then  given by \cite{chta}:
\begin{equation}\label{plane} \phi(x)=\phi(\rho, \Omega)=\cos{\rho}\,
\frac{e^{\pm i(L+1)\rho}}{\sqrt{L+1}}\,  y_{Llm}(\Omega),
\end{equation}
where $y_{Llm}(\Omega)$ are the  hyperspherical harmonics on
$S^3$. It can be  shown
that in the $H =0$ limit and with
\begin{equation}\label{limit}
\rho=Ht,\;\; \alpha=Hr;\;\; HL=k_0=|\vec
k|,\qquad\mbox{with}\quad\theta\,,\varphi\quad\mbox{unchanged}\,,\end{equation}
the functions ($\ref{plane}$) become, when suitably rescaled, the
usual massless spherical waves (with $k^2=(k^0)^2-(\vec k)^2=0$)
\cite{queva}.

With these coordinates, the dS  vector square integrable solutions  formally
read
\begin{equation}
\K_{\mu}(\rho,\Omega)= \frac{\partial x^\alpha}{\partial
X^\mu}\left[\bar Z_{\alpha}- D_{1\alpha}\left(Z\cdot\bar
\partial+H^2 x\cdot Z\right)\right] \cos \rho
\frac{e^{\pm i(L+1)\rho}}{\sqrt{L+1}} y_{Llm}(\Omega)\,.
\end{equation}
We  now discuss the limit $H =0$ in order  to  fix the constant
vector $Z$. The condition is to  recover the minkowskian  four
polarization vectors $\epsilon^{\lambda}_{\mu}(k)$ with
$(\lambda=0,1,2,3)$ (we actually drop the usual parentheses for
$\lambda$ in order to manage handy expressions). These minkowskian
polarization vectors of course satisfy the usual gauge dependent
orthogonality relations \cite{itzu}. In general, the constant
5-vectors $(Z_{\alpha})$ to be selected  will be labelled by
$\lambda=0,1,2,3$ and written $Z^{\lambda}$. This corresponds to
the fact that although expressed as vectors with five components,
the objects in dS space only have four independent components.

Let us at first consider the solutions in the Feynman gauge $c=0$.
Recall that, up to gauge states,  the field solutions in that case
read as
\begin{equation}
\K(x,\xi,Z)=\bar Z \phi+2\,\bar\partial \,Q_0^{-1}\,x\cdot Z
\phi\,.
\end{equation}
A simple  and appropriate choice of $Z$  is then given by
\begin{equation}\label{choice}
Z^{\lambda}_{\alpha}=(\epsilon^{\lambda}_{\mu}(k),
Z^{\lambda}_4=0)\,,
\end{equation}
where $\epsilon^{\lambda}_{\mu}(k)$ are the minkowskian
polarizations which  satisfy the usual relations \cite{itzu}:
\begin{equation}
\epsilon^{0}=n,\;\;\;\;\epsilon^{3}\cdot n=0,\;\;\;\;
\epsilon^{3}\cdot\epsilon^{3}=-1,\;\;\;\;\; n\cdot n=1,\;\;\;
n^0>0,\end{equation}
\begin{equation}   \epsilon^{\lambda}\cdot\epsilon^{\lambda'}=-
\delta_{\lambda \lambda'},\;\;\;\;\;\epsilon^{\lambda}\cdot
n=\epsilon^{\lambda} \cdot k=0,\;\;\;\;\;
\lambda,\lambda'=1,2,\end{equation}
\begin{equation}\eta_{\lambda\lambda'}\epsilon^{\lambda}_{\mu}(k)\epsilon^{*\lambda'}_
{\nu} (k)=\eta_{\mu \nu},
\;\;\;\;\;\epsilon^{\lambda}(k)\cdot\epsilon^{*\lambda'}(k)=\eta
^{\lambda\lambda'}.\end{equation}  This is because qualitatively
one can see using the conformal coordinates and $(\ref{limit})$,
that, in the limit $H=0$, the leading term in $ x\cdot Z\phi$
depends only upon $Z_{4}$. Moreover, $Q_{0}$ and $Q_{0}^{-1}$ do
not modify the $H$ dependence of the functions which they act
upon. For instance $Q_{0}x_{\alpha}=-4x_{\alpha}$. Hence by
setting $Z_{4}$ to zero, one  gets rid of  the logarithmic part of
$(\ref{eq:gauge})$ in the flat limit. Thus we are left with the
polarization vector $\bar{Z}_{\alpha}$ which satisfies
$$
\lim_{H \rightarrow
0}\bar{Z}_{\alpha}=Z_{\mu}=\epsilon^{\lambda}_{\mu}(k)\qquad\mbox{where}
\quad\alpha=0,1,2,3,4\quad\mbox{and}\quad\mu=0,1,2,3.$$

However, for other choices  of gauge ($c\neq0$), the simplest one
 to work with is not given by $(\ref{choice})$. In fact,
similarly to the massive  spin $2$  case \cite{gagata}, we will
impose (at least here for the $c=2/3$ gauge)  $Z^{\lambda}$ to
satisfy
\begin{equation}\label{eq:pola1}
Z^{\lambda}\cdot
Z^{\lambda'}=\eta^{\lambda\lambda'},\quad\eta_{\lambda\lambda'}Z^{\lambda}_{\alpha}Z^{\lambda'}_{\beta}
=\eta_{\alpha\beta}\qquad\forall\;\lambda\,,\lambda'=0,1,2,3\,,
\end{equation}so  that the vector two-point function
assume a maximally symmetric  bi-tensor form of  in  ambient space
notation. This choice presents the great advantage to be
covariantly defined for all components $\alpha$. Its form is not
directly dictated by the flat limit behavior. Rather, we will see
in the following that this choice yields correct expressions on
the level of the two-point function. Indeed, it is easy to show
that the physical polarization vectors
\begin{equation}\label{polaa}
{\cal E }_{2\alpha}^{\lambda}(x,\xi) \stackrel{\mathrm{def}}{=}{\cal E }_{2\alpha}(x,\xi,Z^{\lambda})=2\left(\bar
Z_{\alpha}^{\lambda}-\frac{Z^{\lambda}\cdot x}{x\cdot \xi} \,\bar
\xi_{\alpha}\right)\quad\mbox{with}\quad\xi\cdot{\cal E}_{2\alpha}^{\lambda}(x,\xi)=\xi\cdot Z^{\lambda}=0\,,
\end{equation} satisfy
\begin{eqnarray}
\eta_{\lambda\lambda'}{\cal E}_{2\alpha}^{\lambda}(x,\xi)\,{\cal E}_{2\beta}^{\lambda'}(x,\xi)=
4\left(\theta_{\alpha\beta}-\frac{\bar\xi_{\alpha}\bar\xi_{\beta}}{\left(Hx\cdot\xi\right)^2}\right)\,,
\end{eqnarray}and
\begin{eqnarray}
{\cal E }_{2}^{\lambda}(x,\xi)\cdot{\cal E}_{2}^{\lambda'}(x,\xi)=4 Z^{\lambda}\cdot
Z^{\lambda'}=\eta^{\lambda\lambda'}\,.
\end{eqnarray}
Formally, it is  possible  to compute the  $H=0$ limit of
the latter expressions. Unfortunately,
this particular form ($\ref{polaa}$) is related to the use of the
scalar waves $\left(Hx\cdot\xi\right)^{\sigma}$ with
$\sigma=-1,-2$, and we already pointed out the impossibility to get at the $H=0$ limit nontrivial massless minkowskian
entities.  Note that this isn't true in the massive case where
$\sigma$ can be made $H$ dependent. In that case,  one can
adopt for   $\xi$ a suitable parametrization when one has in view the link
with massive Poincar\'e UIR's:  it is  given by  the orbital
basis
$$\{\xi\;\in C^{+},\xi^{(4)}=1 \}\cup\{\xi\;\in
C^{+},\xi^{(4)}=-1  \},$$ where $\xi$ is defined  in
terms of the four-momentum $(k^0, \vec{k})$ of a minkowskian
particle (for details see \cite{gagata,eric})
\begin{equation}
\xi_{\pm}=\left(\frac{k^0}{mc}=\sqrt{\frac{\vec
k^2}{m^2c^2}+1},\frac{\vec k}{mc},\pm 1\right).
\end{equation}
To summarize, one could say that in the ambient space formalism
the massless polarization vectors in dS space look very similar
 to their minkowskian counterparts. Unfortunately, the
corresponding $H=0$ limit yields expressions which aren't easy to
interpret from a minkowskian point of view.

Let us end this part with a remark concerning the gauge states.
We have said that any solution to Eq. $(\ref{eq:wave1})$ is
defined up to an arbitrary gauge field $D_{1}\phi_{M}$. As a
matter of fact we already had a gauge solution in the expression   $\K_{2\alpha}(x)$ through ${\cal E}_{2}(x,Z,\xi)$ as
$$ \K_{2\alpha}(x,\xi,Z)= \cdots -6 \frac{Z\cdot \xi}{(H x\cdot\xi)^4}\,\bar
\xi_{\alpha}\,.$$ Here one recognizes $2\, Z\cdot\xi
D_{1}\phi_{M}$ with $\phi_{M}=\left(Hx\cdot\xi\right)^{-3}$. Since
the gauge  solutions are present in $\K_{2\alpha}(x)$, we
can choose, without any loss of generality, to work with the
solutions
\begin{equation}\label{sol1}
\K_{1\alpha}(x,\xi,Z)={\cal
E}_{1\alpha}(x,\xi,Z)(Hx\cdot\xi)^{-1}\,,\qquad\qquad
\K_{2\alpha}(x,\xi,Z)={\cal
E}_{2\alpha}(x,\xi,Z)(Hx\cdot\xi)^{-2}\,,
\end{equation}
where
\begin{align*}
\label{}
    {\cal E}_{1\alpha}(x,\xi,Z)=&\bar Z_{\alpha}- 2 \frac{Z\cdot \xi}{(H
x\cdot\xi)^2} \,\bar \xi_{\alpha}, \\
      {\cal E}_{2\alpha}(x,\xi,Z)=& 2\bar Z_{\alpha}-2\frac{Z\cdot x}{x\cdot \xi}
\,\bar \xi_{\alpha}- 6 \frac{Z\cdot \xi}{(H x\cdot\xi)^2} \bar
\xi_{\alpha}.
\end{align*}

\subsection{Analytic vector waves}

The vectors  $\K(x)$ given by formula (\ref{sol}) are not globally
defined since the waves $(Hx\cdot\xi)^{\sigma}$ are singular on
three dimensional light-like manifolds \cite{brmo}. For  a global
definition, they  have to be viewed as distributions \cite{gesh}.
More precisely,  we will consider the  boundary values of analytic
continuations of the solutions $\K(x)$ to tubular domains in the
complexified de Sitter space $X_H^{(c)}$. The complexified dS
space is defined by:
\begin{eqnarray}
X_H^{(c)}&=&\{z=x+iy\in  \setC^5;\;\;\eta_{\alpha \beta} z^\alpha
z^\beta=(z^0)^2-\vec z\cdot\vec
z-(z^4)^2=-H^{-2}\}\nonumber\\
&=&\{ (x,y)\in\setR^5\times \setR^5;\;\; x^2-y^2=-H^{-2},\; x\cdot
y=0\}\,.\nonumber
\end{eqnarray}
For a generic $\sigma$ and an univalued determination of the expression
$\left(z\cdot\xi\right)^{\sigma}$ we adopt the principal
determination in
$$\left(z\cdot\xi\right)^{\sigma}=\exp\left(\sigma\left[\log\vert z\cdot\xi\vert+i\mbox{arg}(z\cdot\xi)\right]\right)\quad\mbox{with}\quad\mbox{arg}(z\cdot\xi)\in\;]-\pi,\pi[\,,$$
and   characterize $z$ so that we fix the sign of the
imaginary part of $\left(z\cdot\xi\right)$. Let us introduce the
forward and backward tubes of $X_H^{(c)}$. First of all, let
$T^\pm=\setR^5-iV^\pm$ be the forward and backward tubes in
$\setC^5$. The domain $V^+$(resp. $V^-)$ stems from the causal
structure on $X_H$:
\begin{equation}
V^\pm=\{ x\in \setR^5;\;\; x^0\stackrel{>}{<}\sqrt{\parallel \vec
x\parallel^2+(x^4)^2} \}. \label{cone}
\end{equation}
We then introduce their respective intersections with $X_H^{(c)}$,
\begin{equation}
{\cal T}^\pm=T^\pm\cap X_H^{(c)},
\end{equation}
which are the tubes of $X_H^{(c)}$. Finally we define the
``tuboid'' above $X_H^{(c)}\times X_H^{(c)}$ by
\begin{equation}
{\cal T}_{12}=\{ (z,z');\;\; z\in {\cal T}^+,z'\in {\cal T}^- \}.
\end{equation}
Details are given in \cite{brmo}. When $z$
varies in ${\cal T}^+$ (or ${\cal T}^-$) and $\xi$ lies in the
positive cone ${\cal C}^+$ the  wave solutions are globally
defined because the imaginary part of $(z\cdot \xi)$ has a fixed
sign and $z\cdot\xi\neq 0$.

Now, if $\sigma = -1, -2$, the wave solutions are univalued, of course,
but still singular, and an analytical continuation is still needed in order
to view them as well-defined objects. Therefore, we define the  de Sitter tensor wave
${\K}_{\alpha}(x)$ as the boundary value of the analytic
continuation to the future tube of Eq. (\ref{sol}). Hence, for $z
\in {\cal T}^+ $ and $\xi \in {\cal C}^+$ one gets the two
solutions
\begin{equation}\label{eq:dswave}
{K}_{1\alpha}(z)={\cal E}_{1\alpha}^{\lambda}(z,\xi) \left(
Hz\cdot \xi\right)^{-1},\quad\mbox{and}\quad
{K}_{2\alpha}^{*}(z^{*})={\cal E}_{2\alpha}^{*\lambda}(z^{*},\xi)
\left( Hz\cdot \xi\right)^{-2}.
\end{equation}
The corresponding boundary values are
\begin{eqnarray}
\mbox{bv}\;{K}_{1\alpha}(z)&\equiv& {\K}_{1\alpha}(x)={\cal
E}_{1\alpha}^{\lambda}(x,\xi) \left(
H(x+i\epsilon)\cdot \xi\right)^{-1}\,,\nonumber\\
\mbox{bv}\;{K}_{2\alpha}(z)&\equiv&{\K}_{2\alpha}(x)={\cal
E}_{2\alpha}^{\lambda}(x,\xi) \left( H(x+i\epsilon)\cdot
\xi\right)^{-2}\,,
\end{eqnarray}
with $\epsilon\,\in\,V^{+}$ arbitrarily small.

 \setcounter{equation}{0}

\section{The two-point function}

In a previous work concerning the massive vector case \cite{gata},
we have constructed the field theory from the Wightman two-point
function. The two-point function had to satisfy the conditions of
 a) positivity, b) locality, c) covariance, d) normal
analyticity, e) transversality and d) divergencelessness in order
to properly encode the theory of free fields on dS space. In the
``massless"  case, the divergencelessness condition cannot be
maintained if one wishes to preserve the covariance condition.
Consequently the field equation is solved in a larger $c$-dependent
space endowed with an indefinite inner product. On the level of
the two-point function this forces to abandon the positivity. The
corresponding two-point function cannot be covariantly  separated
into a positive (physical) and a negative part.

Given the solutions $(\ref{sol1})$ one defines the analytic
two-point function  explicitly in terms of the following class of
integral representations
\begin{equation}\label{eq:tpfunction}
W _{\alpha \alpha'}(z,z') =a_0 \int_{\gamma} (H z\cdot \xi)^{-1}
(H z'\cdot\xi)^{-2}\,\eta_{\lambda\lambda'}{\cal
E}^{\lambda}_{1\alpha }(z,\xi) \,{\cal E}^{*\lambda'}_{2\alpha'}
({z'}^{*},\xi)\,d\sigma_{\gamma}(\xi)\,,
\end{equation}
where  $d\sigma_{\gamma}(\xi)$ is the natural ${\cal C}^{+}$
invariant measure on $\gamma$, induced from the $\setR^{5}$
Lebesgue measure \cite{brmo} and the normalization constant $a_0$
is fixed by local Hadamard condition. The tensor \begin{equation}
\label{tensorpol} \eta_{\lambda\lambda'}{\cal
E}^{\lambda}_{1\alpha }(z,\xi) {\cal E}^{\lambda'}_{2\alpha'}
(z',\xi)\equiv T_{\alpha \alpha'}(z,z',\xi)
\end{equation} is a covariant homogeneous bi-vector, in ambient space notation, of degree $0$ in the variables
$z,z'$ and $\xi$.  Of course, this choice is not unique but it is
motivated by several facts. First, in the massive vector case, we
have also constructed the two-point function  with  two vectors
based on the product $(Hz'\cdot\xi )^{\sigma}(Hz\cdot\xi
)^{-\sigma-3}$ (see Reference\cite{gata}). Moreover, in the
conformally scalar case this type of two-point function coincides
with the expression found for that field in \cite{tag} as we will
show below. Finally,   given the set of modes
${\K}_{1\alpha}(z),{\K}_{2\alpha}(z)$, we will actually show that
the only simple product  yielding  a causal two-point function is
the one presented in $(\ref{eq:tpfunction})$. The boundary value
of (\ref{eq:tpfunction}) defines the two-point function in terms
of global plane waves on $X_H$.

The analytic two-point function (\ref{eq:tpfunction}) can be
expressed in terms of an analytic  scalar two-point function
without resorting to any explicit calculation of the integral.
Actually, following  Allen and Jacobson in Reference \cite{allen1}, we
will write the two-point functions in de Sitter space
 in terms of bivectors. These are functions
of two points $(x,x')$ which behave like vectors under coordinate
transformations at either point. The bivectors are called
maximally symmetric if they respect the de Sitter invariance. As
shown in  \cite{gagata} and also proved in Appendix \ref{appC},
any maximally symmetric bivector can be expressed in ambient space
notations as a sum of the two basic bivectors
$\theta_{\alpha}\cdot\theta'_{\alpha' }$ and $ \bar
\partial_{\alpha} \bar\partial'_{\alpha'}$. Thus we can write
\begin{equation}\label{eq:tpf3}
 W_{\alpha \alpha'}(z,z')=\theta_{\alpha}\cdot\theta'_{\alpha' }
W_0(z,z')+H^{-2}\bar \partial_{\alpha} \bar
\partial'_{\alpha'}W_1(z,z')\,.
\end{equation}
Of course, it will be verified that this expression coincides with
the two-point function constructed from the modes (\ref{sol1}) in
formula (\ref{eq:tpfunction}) (in the case $c=2/3$). By imposing
the bivector (\ref{eq:tpf3}) to obey Eq. (\ref{eq:wave1}) in
variable $z$ or $z'$, one finds from (\ref{eq:relation1}) the
following relations
\begin{equation}\label{syst}
\left(Q_0-2\right)W_0(z,z')=0\,,\quad
Q_0\bar\partial'W_{1}(z,z')=2H^2 z\cdot\theta'W_{0}(z,z')-c\,\bar\partial\cdot
W_{}(z,z').
\end{equation}
Let us first examine $W_{0}(z,z')$. This analytic two-point
function corresponds to the massless conformally coupled scalar
field associated with the complementary series of unitary
representations \cite{brmo,ta}. The Wightman scalar two-point
function  ${\cal W}_{0}(x,x')$  in that case is  given by
\cite{brmo}
\begin{equation}\label{eq:sca}
{\cal W}_{0}(x,x')=\mbox{bv}\;W_{0}(z,z')\qquad\mbox{with}\qquad
W_{0}(z,z')=\, c_0 \int_{\gamma} (H z\cdot \xi)^{-1}
(Hz'\cdot\xi)^{-2}d\sigma_{\gamma}(\xi) \,.
\end{equation}
The normalization constant $c_0$ is determined by imposing the
Hadamard condition on the two-point function. This has been done
in Ref. \cite{brmo} where the scalar two-point function has been
rewritten in terms of the generalized Legendre function  for well-chosen
points $z,z'\,\in{\cal T}_{12}^{}$ in the domain defined
by $(z-z')^2<0$. For instance
$z=(-iH^{-1}\cosh\varphi,-iH^{-1}\sinh\varphi,0,0,0)$ and
$z'=(iH^{-1},0,0,0,0)$ with $z\cdot z'=H^{-2}\cosh\varphi$. It has
been established that
\begin{equation}\label{eq:tpf2}
W_{0}(z,z') = C_0
P_{-1}^{(5)}(-\z)=\frac{-H^2}{8\pi^2}\frac{1}{1-{\cal Z }(z,z')},
\;\;
\end{equation}  with
$\z=-H^{2}z\cdot z'$,  $C_{0}=-2i\pi^2  c_0$ and
\begin{equation}
c_0=\frac{iH^2}{2^5\pi^4 }. \label{eq:nor}
\end{equation}
For details about the normalization see \cite{brmo,eric} and for
the actual explicit form of $W_{0}(z,z')$ in terms of $\z$, see
\cite{ta}. The function
$P_{\sigma}^{(5)}(\cosh\varphi)=P_{\sigma}^{(5)}(-\z)$ is related
to the associated Legendre function
${\mathfrak{P}}_{\sigma+1}^{-1}(\cosh \varphi ) $ through
$$P_{\sigma}^{(5)}(\cosh\varphi) =-4i\left(\sinh
\varphi\right)^{-1}{\mathfrak{P}}_{\sigma+1}^{-1}(\cosh\varphi).$$
The boundary value of Eq. (\ref{eq:tpf2}) yields \cite{ta}
\begin{equation}\label{scacon}
{\cal W}_0(x,x')=\frac{-H^2}{8\pi^2}\left[P\frac{1}{1-{\cal
Z}(x,x')}-i\pi\epsilon(x^0-{x'}^0)\delta(1-{\cal Z}(x,x'))\right].
\end{equation}
where $P$ denotes the principal part and
$\epsilon(x^0-{x'}^0)=1,0,-1$ whether  one has
$(x^0-{x'}^0)>,=,$ or $<0$ respectively. This is exactly the two-point function
of the conformally coupled scalar field given in  \cite{tag}.

We now consider the second relation in $(\ref{syst})$. Since the
divergence of $W(z,z')$ reads as
$$\bar\partial \cdot W(z,z')=4H^2 z\cdot\theta' W_0(z,z')+\theta'\cdot\bar\partial
W_0(z,z')-Q_0\bar\partial' W_1(z,z')\,,$$ one  easily rewrite
the second relation in  $(\ref{syst})$ as
\begin{equation}\label{eq:wg}
Q_0 \bar \partial'{ W}_1(z,z')=-\frac{c}{1-c}\left[\theta'
\cdot\bar
\partial +H^2z\cdot
\theta'\right]
W_0(z,z')+\frac{2-3c}{1-c}H^2z\cdot\theta'W_0(z,z').
\end{equation}

\subsection{\bf The minimal case, $c=2/3$}

Again, we first consider  the simple case $c=\frac{2}{3}$. The
above equation  simplifies to
\begin{equation}
Q_0 \bar
\partial' W_1(z,z')=-2\left[\theta'\cdot \bar
\partial  +H^2z\cdot\theta'\right] W_0(z,z')\,,
\end{equation}
which is satisfied if
\begin{equation}
\bar \partial' W_1(z,z')=-\left[ \theta'\cdot\bar
\partial+H^2z\cdot\theta'\right]W_0(z,z').
\end{equation}
Thus, we can write the analytic two-point function in the form
\begin{equation}
 W_{\alpha \alpha'}(z,z')=D_{\alpha \alpha'}
W_0(z,z'),\end{equation} where
\begin{equation}\label{DAA}
D_{\alpha \alpha'}=\theta_{\alpha
}\cdot\theta'_{\alpha' }-H^{-2}\bar
\partial_{\alpha} \left[\theta'_{\alpha' }\cdot\bar \partial
+H^2z\cdot\theta'_{\alpha' }\right].
\end{equation}
The  vector two-point function can be developed to
\begin{equation}
W_{\alpha \alpha'}(z,z')=-H^{-2}\bar \partial_{\alpha}\,
\theta'_{\alpha'}\cdot\bar
\partial\, W_0(z,z')-x\cdot\theta'_{\alpha'} \bar
\partial_{\alpha}W_0(z,z')\,,
\end{equation}
and by simple derivation  one finally obtains
\begin{equation}
W_{\alpha \alpha'}(z,z') = c_0 \int_{\gamma}\left(
\theta_{\alpha}\cdot\theta'_{\alpha'}-\frac{2\,\bar\xi'_{\alpha'}\bar\xi_{\alpha}}{(H
z\cdot\xi)^2}\right) (H z\cdot \xi)^{-1}
(Hz'\cdot\xi)^{-2}d\sigma_{\gamma}(\xi).
\end{equation}
In terms of the scalar two-point function, the vector two-point
function can be written in the following way
$$W_{\alpha \alpha'}(z,z')=-\theta'_{\alpha'}\cdot\theta_\alpha\,\z\dz
W_0(z,z') +H^2(\theta'\cdot z)(\theta\cdot z')\left(
2\frac{d}{d{\cal Z}}+{\cal Z}\frac{d^2}{d{\cal
Z}^2}\right)W_0(z,z'),$$ where we have used
\begin{equation}
\bar\partial_\alpha W_0(z,z')=-H^2z'\cdot\theta_\alpha\frac{d
}{d{\cal Z}}W_0(z,z').
\end{equation}
It is explicitly  shown in Appendix $D$ that this two-point
function agrees with Eq. (\ref{eq:tpfunction}) when the constant
vectors $Z^{\lambda}$ satisfy
$\eta_{\lambda\lambda'}Z_{\alpha}^{\lambda}Z_{\beta}^{\lambda'}=\eta_{\alpha\beta}$.
Taking a closer look at the two-point function with a
$T_{\alpha\alpha'}$ introduced in Eq. (\ref{tensorpol}) given by
\begin{equation}
T_{\alpha\alpha'}(z,z',\xi)=\theta_{\alpha}\cdot\theta'_{\alpha'}-\frac{2\,\bar\xi'_{\alpha'}\bar\xi_{\alpha}}{(H
z\cdot\xi)^2}\,,
\end{equation}
one notices that it is  analogous to   the minkowskian gauge-dependent
polarization sum which in general reads :
\begin{equation}\label{eq:propa}
T_{\mu\nu}(k)=\eta_{\mu\nu}-\frac{c}{1-c}\frac{\,k_{\mu}k_{\nu}}{k^2}\,,
\end{equation}
with $c/(1-c)=2$ for $c=2/3$.

Now, taking the boundary value limit, it can be proved, by using the
same methods as in \cite{gata}, that the two-point function ${\cal
W}_{\alpha \alpha'}(x,x') = W_{\alpha \alpha'}(z,z')\mathrm{bv}\, $
satisfies the following conditions:
\begin{enumerate}
\item[a)] {\bf Indefinite sesquilinear form}

For any test function $f_\alpha \in {\cal D}(X_H)$, we have an
indefinite sesquilinear form that is defined by
\begin{equation} \int _{X_H \times X_H}
f^{*\alpha}(x){\cal W}_{\alpha \alpha'}(x,x')
f^{\alpha'}(x')d\sigma(x)d\sigma(x'),\end{equation} where $ f^*$
is the complex conjugate of $f$ and $d\sigma (x)$ denotes the
dS-invariant measure on $X_H$ \cite{brmo}. ${\cal D}(X_H)$ is the
space of  $C^\infty$ functions with compact support in $X_H$.

\item[b)] {\bf Covariance}

The two-point function satisfies the covariance property
\begin{equation}g^{-1}{\cal W} (g x,g x')g={\cal
W}(x,x').
\end{equation}
where $g\in SO_0(1,4)$.

Indeed, let us first write   the group action on the dS modes.
From Eq.(\ref{solution}), we recall that the latter  are given by
\begin{equation}
\K_{\alpha}(x,\xi,Z)={\cal
E}_{\alpha}(x,\xi,Z)\,\phi(x)+HD_1\phi_M\,,
\end{equation}
with
\begin{equation}
{\cal E}_{\alpha}(x,\xi,Z)=
 \left[-\sigma\bar
Z_{\alpha}-\sigma(\sigma+1)\frac{Z\cdot x}{x\cdot\xi} \,  \bar
\xi_{\alpha}- \sigma(\sigma-1)\frac{Z\cdot\xi} {(H
x\cdot\xi)^2}\,\bar \xi_{\alpha}\right].
\end{equation}
Now, one easily shows that (also recall
$\phi_{M}=\left(Hx\cdot\xi\right)^{\nu}$ with $\nu=0,-3$)
\begin{equation}
{\cal E}_{\alpha}(g^{-1}x,\xi,Z)=(g^{-1})_{\alpha}^{\delta}{\cal
E}_{\delta}(x,g\xi,gZ)\,\quad\mbox{and}\quad
D_{1\alpha}\left(Hg^{-1}x\cdot\xi\right)^{\nu}=(g^{-1})_{\alpha}^{\delta}D_{1\delta}
\left(Hx\cdot g\xi\right)^{\nu}\,.
\end{equation}
Therefore the group action on the dS modes reads :
\begin{equation}\label{eq:graction}
\left(U(g) \,\K\right)_{\alpha}(x,\xi,Z) = g_{\alpha}^{\gamma}
\K_{\gamma }(g^{-1} x,\xi,Z)=\,\K_{\alpha}(x,g\xi,gZ).
\end{equation}
The simplicity of the group action again shows the efficiency of
the ambient space formalism. Finally since the integral
$(\ref{eq:tpfunction})$ is independent of a specific choice of $\xi$
(orbital basis) or $Z$ this proves the covariance property.

\item[c)] {\bf Locality}

For every space-like separated pair $(x,x')$, {\it i.e.} $x\cdot
x'>-H^{-2}$,
\begin{equation}{\cal W}_{\alpha \alpha'}(x,x')={\cal
W}_{ \alpha' \alpha}(x',x).
\end{equation}

In order to prove this locality condition, we use the identity
$$ W^{*}_{\alpha\alpha'
}(z^*,z^{\prime*})=W_{\alpha\alpha' }(z',z)$$ easily checked using
$(\ref{eq:tpfunction})$  and the following relation
\begin{equation}\label{hermiticity}
 W_{\alpha
\alpha'}(z,z')=W_{\alpha'\alpha}^*(z^{\prime*},z^*).
\end{equation}
The latter is valid for space-like separated points
$z,z'$ (satisfying $(z-z')^2<0$ ) since it is  based on the fact
that, for space-like separated points $z,z'$, one can rewrite the
two-point function  as :
$$W_{\alpha\alpha'}(z,z')=D_{\alpha\alpha'}(z,z')P_{\sigma}^{(5)}(-\z)\quad\mbox{with}\quad
D_{}^{*}(z^{\prime*},z^{*})=D_{}(z',z)\,.$$  Then we use the
relations
$$P_{-2}^{(5)}(-\z) =-4i\left(\sinh \varphi\right)^{-1}{\mathfrak{P}}_{-1}^{-1}(\cosh\varphi)
=-4i\left(\sinh
\varphi\right)^{-1}{\mathfrak{P}}_{0}^{-1}(\cosh\varphi)=P_{-1}^{(5)}(-\z)\,,
$$ which follow from the Legendre function property  ${\mathfrak{P}}_{\nu}^{\mu}(x)={\mathfrak{P}}_{-\nu-1}^{\mu}(x)$.
One finally gets
$$W_{\alpha\alpha'
}(z,z')=W_{\alpha'\alpha}^*(z^{\prime*},z^*)=W_{\alpha'
\alpha}(z',z).
$$
Finally, the space-like separated pair ($x,x'$) lies in the same
orbit of the complex dS group as the pairs ($z,z'$) and
($z'^{*},z^*$). Therefore the locality condition ${\cal W}_{\alpha
\alpha'}(x,x')={\cal W}_{\alpha' \alpha}(x',x)$ holds for the
space-like separated points $x,x'$.

\item[d)] {\bf Normal analyticity}

${\cal W}_{\alpha \alpha' }(x,x')$ is the boundary value (in the
sense of distributions) of an analytic function $W_{\alpha
\alpha'}(z,z')$. The analyticity properties of the tensor
Wightman two-point function in the tuboid $ {\cal T}_{12}=\{
(z,z');\;\; z\in {\cal T}^+,z'\in {\cal T}^- \} $ follow from the
analyticity properties of the dS tensor waves ($\ref{eq:dswave}$).
\item[e)] {\bf Transversality}
\begin{equation} x\cdot {\cal W}(x,x')=0=x'\cdot {\cal
W}(x,x'),\end{equation} The  transversality with respect to $x$
and $x'$ is guaranteed since the dS modes  $\K(x)$ are transverse
by construction.
\end{enumerate}

\subsection{\bf The general case $c\neq 2/3$}

Let us briefly  consider the case $c\neq 2/3$. We can write the
equation (\ref{eq:wg}) in the form
\begin{equation}
 \bar \partial' W_1(z,z')=-\frac{c}{2(1-c)}\left[\theta'\cdot\bar
\partial  +H^2z\cdot\theta'\right]
W_0(z,z')+\frac{2-3c}{1-c}H^2 Q_0^{-1}z\cdot\theta'
W_0(z,z'),\end{equation} and  the vector two-point function
becomes
\begin{align}
\nonumber W_{\alpha \alpha'}(z,z')&=\theta_{\alpha }\cdot\theta'_{\alpha' }
W_0(z,z')-\frac{c}{2(1-c)}H^{-2}\bar
\partial_{\alpha}\left[\theta'_{\alpha' }\cdot\bar
\partial  +H^2z\cdot\theta'_{\alpha'
}\right]W_0 (z,z')\\
& +\frac{2-3c}{1-c}\bar
\partial_{\alpha} Q_0^{-1}z\cdot\theta'_{\alpha' }W_0(z,z').
\end{align}
In order to distinguish the specific value  $c = 2/3$, a convenient form of the above expression is
\begin{equation}
{W}_{\alpha \alpha'}^c=W_{\alpha
\alpha'}^{\frac{2}{3}}+\frac{\frac{2}{3}-c}{H^2(1-c)}\bar
\partial_{\alpha}
Q_0^{-1}\bar\partial\cdot W_{\alpha'}^{\frac{2}{3}},
\end{equation}
where the $W_{\alpha \alpha'}^{\frac{2}{3}}$ is the two-point
function corresponding to  $c=2/3$. The singularity appears in the
term $ \bar\partial Q_0^{-1}\bar\partial\cdot
W_{\alpha'}^{\frac{2}{3}}$ of the above equation.

\setcounter{equation}{0}

\section{The quantum field}

Let us now write the field corresponding to our two-point
function. For any test function $f_{\alpha}\in {\cal D}(X_H)$, we
define the vector-valued distributions taking values in the space
generated by the modes $\K_{\alpha}(x,\xi)\equiv
\mbox{bv}\, K_{\alpha}(z,\xi)$ by:
\begin{equation}
x\rightarrow p_{1\alpha}(f)(x)=\sum_{\lambda=0}^{3}
\zeta_{\lambda} \int_{\gamma}d\sigma_{\gamma}(\xi)
{\K}^{\lambda}_{2\xi}(f)\,{\K}_{1\alpha}^{\lambda}(x,\xi)\,,
\end{equation}
and
\begin{equation}
x\rightarrow p_{2\alpha}(f)(x)=\sum_{\lambda=0}^{3}\zeta_{\lambda}
\int_{\gamma}d\sigma_{\gamma}(\xi)
{\K}^{\lambda}_{1\xi}(f)\,{\K}_{2\alpha}^{\lambda}(x,\xi) \,,
\end{equation}
with $\zeta_{0}=+1$ and $\zeta_{\lambda}=-1$ for $\lambda=1,2,3$
and  where $\K^{\lambda}_{n\xi}(f)$ with n$=1,2$ is the smeared
form of the modes:
\begin{equation}
{\K}^{\lambda}_{n\xi}(f)=\int_{X_{H}}{\K}_{n\alpha}^{*\lambda}(x,\xi)f^{\alpha}(x)d\sigma(x)\,.
\end{equation}
The space generated by the $p(f)$'s is equipped with the
indefinite invariant inner product
\begin{equation}
\langle p(f),p(g)\rangle=\int_{X_{H}\times X_{H}}
f^{*\alpha}(x){\cal W}_{\alpha\alpha'}(x,x')
g^{\alpha'}(x')d\sigma(x')d\sigma(x)\,.
\end{equation}
As usual, one could be attempted  to define the fields as  operator-valued
distributions,
\begin{equation}
{\cal
K}(f)=a\left(p(f)\right)+a^{\dagger}\left(p(f)\right)\qquad\mbox{with}\qquad
p(f)=p_{1}(f)+p_{2}(f)\,,
\end{equation}
where the operators $a(\K^{\lambda}(\xi)) \equiv a^{\lambda}(\xi)$
and $a^{\dagger}(\K^{\lambda}(\xi)) \equiv
a^{\dagger\lambda}(\xi)$ are respectively antilinear and linear in
their arguments. One would get the hermitian field:
\begin{equation}
{\cal K}(f)=\sum_{\lambda=0}^{3}\zeta_{\lambda}\int_{\gamma}
d\sigma_{\gamma}(\xi)\left[\K^{*\lambda}_{\xi}(f)\,a^{\lambda}(\xi)
+\K^{\lambda}_{\xi}(f)\,a^{\dagger\lambda}(\xi)\right]\,,
\qquad\mbox{with}\qquad \K^{\lambda}_{\xi}(f)=\sum_{n=1}^{2} {\K}^{\lambda}_{n\xi}(f) .
\end{equation}
The unsmeared operator would read
\begin{equation}\label{eq:field}
{\cal K}_{\alpha}(x)=
\sum_{\lambda=0}^{3}\zeta_{\lambda}\int_{\gamma}
d\sigma_{\gamma}(\xi) \left[ \;{\cal
\K}_{\alpha}^{\lambda}(x,\xi)\, a^{\lambda}(\xi) +{\cal
\K}_{\alpha}^{*\lambda} (x,\xi)\,
a^{\dagger\lambda}(\xi)\right]\,,
\end{equation}
where $a^{\lambda}(\xi)$ satisfies the canonical commutation
relations (ccr) and is defined by
$$a^{\lambda}(\xi)|\Omega>=0.$$
The field equation (\ref{eq:field}), however, is problematic,
since the intergral involved does not have a unique solution due
to its degrees of homogeneity. The homogeneity degrees of
$a^{\lambda}(\xi)$ are not fixed, {\it i.e.}, for
${\K}^{\lambda}_{1\xi}$ mode the degree is $-2$ and for
${\K}^{\lambda}_{2\xi}$ mode is $-1$.

In order to set aside this problem, a causal field, which is
constructed from the modes (\ref{eq:dswave}), {\it i.e.}
\begin{equation}\label{eq:lebonchamp}
{\cal K}(f)=a\left(p_1(f)\right)+a^{\dagger}\left(p_2(f)\right)\,,
\end{equation} should replaced (VI.5). It is clear that this field is not hermitian.

The measure satisfies
$d\sigma_{\gamma}(l\xi)=l^{3}d\sigma_{\gamma}(\xi)$ and the field
operator homogeneity is :${\K}^{\lambda}_{n\alpha}(x,l
\xi)=l^{-n}{\K}^{\lambda}_{n\alpha}(x,\xi)$ . This leads to the
homogeneity condition
$$ \;a^{\lambda}(l \xi)\equiv a(\K_{n}^{\lambda}(l\xi))=a(l^{-n}\K^{\lambda}(\xi))
=l ^{-n}a^{\lambda}(\xi).$$ The integral representation
(\ref{eq:field}) is independent of the orbital basis $\gamma$ as
explained in \cite{brmo}.  For the hyperbolic type submanifold
$\gamma_{4}$ given by
$$\gamma_{4}=\{\xi\;\in C^{+},\xi^{(4)}=1 \}\cup\{\xi\;\in
C^{+},\xi^{(4)}=-1  \},$$ the measure is
$d\sigma_{\gamma_{4}}(\xi)=d^{3}\vec{\xi}/\xi_{0}$ and the ccr's are
represented by

\begin{equation}
[a^{\lambda}(\xi),a^{\dagger\lambda'}(\xi')]=
\eta^{\lambda\lambda'}\xi^0\delta^3(\vec\xi-\vec\xi').
\end{equation}
The field commutation relations are
\begin{equation}
\left[{\cal K}_{\alpha}(x),{\cal K}_{\alpha'}(x')\right]=2i\mbox{Im}
 \,\langle p_{\alpha}(x),p_{\alpha'}(x')\rangle=2i
\mbox{Im}\,{\cal W}_{\alpha\alpha'}(x,x')=2i D_{\alpha\alpha'}
\mbox{Im}{\cal W}_0(x,x')\,,
\end{equation}
where  ${\cal W}_{\alpha\alpha'}(x,x')$ and ${\cal W}_0(x,x')$ are
respectively the vector two-point function  and the conformally
coupled scalar field two-point function \cite{tag,ta,brmo}. Using
formula $(\ref{scacon})$ one has
\begin{equation}
\mbox{Im}{\cal W}_0(x,x')=\frac{H^2}{8\pi}\, \epsilon
(x^0-{x'}^0)\, \delta ({\cal Z}(x,x')-1)\,\,,
\end{equation}
where we have used  ${\cal Z}(x,x')=-H^2 x\cdot x'=1+\frac{H^2}{2}
(x-x')^2 \equiv \cosh H \sigma (x,x') $ and
\begin{equation} \epsilon (x^0-{x'}^0)=\left\{\begin{array}{clcr}
1&x^0>{x'}^0 \\
0&x^0={x'}^0\\   -1&x^0<{x'}^0.\\   \end{array}
\right.\end{equation} Finally one obtains the commutator :
\begin{equation} \label{proplc}
iG_{\alpha\alpha'}(x,x')=[K_\alpha(x),
K_{\alpha'}(x')]=\frac{iH^2}{4\pi}D_{\alpha\alpha'}\epsilon
(x^0-x^{\prime 0})\, \delta ({\cal Z}(x,x')-1),
\end{equation}
with $D_{\alpha\alpha'}$ given by (\ref{DAA}). Light cone
propagation of the vector field is apparent in the r.h.s. of (\ref{proplc}).

\section{Conclusion}
We have quantized the massless vector field in de Sitter space-time
by adapting to this specific situation the content of previous
works: ambient space formalism, construction of modes, de Sitter
covariance and Gupta-Bleuler triplets, construction of the Wightman
two-point function, and eventually covariant quantization of the
field. The next step will be to examine the possibility of
construction of a dS covariant QED, based on the present work and on
the explicit construction of dS ``massive'' Dirac fields which has
been carried out in \cite{bagamota}. Of course, the main questions
will pertain to the physical interpretation of such a formalism,
since Physics in de Sitter space-time is far from being a clear and
familiar domain of investigation.
\newpage
\setcounter{equation}{0}
\setcounter{section}{0}

\appendix

\section{The unitary irreducible representations of SO(1,4)}\label{appA}
The UIR's may be labelled by
a pair of parameters $\Delta =(p,q)$ with $2p \in {\setN}$ and $q
\in {\setC}$, in terms of which the eigenvalues of $Q^{(1)}$ and
$Q^{(2)}$ are expressed as follows \cite{dix,tak}:
$$
{Q^{(1)}}=[-p(p+1)-(q+1)(q-2)]{\1}, \quad
{Q^{(2)}}=[-p(p+1)q(q-1)]{\1}.
$$
According to the possible values for $p$ and $q$, three series of
inequivalent unitary representations may be distinguished: the
principal, complementary and discrete series.

\vspace{0.3cm} \noindent {\Large\color{red} The Principal series of
representations :}\vspace{0.2cm}

Also called ``massive'' representations, they are  denoted  by
$U_{p,\nu}$, and labelled with \quad $\Delta=(p,q)=\left(p,{1\over
2}+i\nu\right)$ where
\begin{eqnarray}
&&p=0,1,2,\dots \quad {\rm and} \quad  \nu\geq 0 \quad {\rm
or},\nonumber\\
&&p={1\over 2},{3\over 2},\dots \quad \ \, {\rm and} \quad  \nu
>0.\nonumber
\end{eqnarray}
The operators $Q^{(1)}$ and $Q^{(2)}$ are fixed  respectively to the
following values:
$${Q^{(1)}} =\left[ \left( {9\over 4}+{\nu ^2} \right)-p(p+1) \right]\,
{\1},\qquad {Q^{(2)}} =\left[ \left( {1\over 4}+{\nu ^2}
\right)p(p+1) \right]\, {\1}.$$

\vspace{0.3cm}
\noindent
{\Large\color{blue} The  complementary series representations :}\vspace{0.2cm}

The  complementary series is denoted by  $V_{p,\nu}$  with
$\Delta=(p,q)=(p,{1\over 2}+\nu)$ and
\begin{eqnarray}
&&p=0 \quad {\rm and} \quad  \nu\in \setR\ ,\ 0<|\nu|<{3\over 2}
\quad {\rm
or},\nonumber\\
&&p=1,2,3,\dots \quad {\rm and} \quad \nu\in{\setR}\ ,\
0<|\nu|<{1\over 2}.\nonumber
\end{eqnarray}
The operators
$Q^{(1)}$ and $Q^{(2)}$ assume the following values
$$
{Q^{(1)}} =\Bigl[ \bigl( {9\over 4}-{\nu ^2} \bigr)-p(p+1)
\Bigr]\,{\1}, \qquad{Q^{(2)}} =\Bigl[ \bigl( {1\over 4}-{\nu ^2}
\bigl)p(p+1) \Bigr]\, {\1}\,.$$

\vspace{0.3cm}
\noindent
{\Large\bf The discrete series of representations }\vspace{0.2cm}

The elements of the  discrete series of representations are
denoted by $\Pi_{p,0}$ and $\Pi^{\pm}_{p,q}$ where the signs $\pm$ would stand for helicity in the massless cases.
The relevant values for the couple  $\Delta=(p,q)$ are
\begin{eqnarray}
&&p=1,2,3,\dots  \quad {\rm and} \quad q=p,p-1,\dots,1,0 \quad {\rm or},\nonumber\\
&&p={1\over 2},{3\over 2},\dots \quad {\rm and} \quad
q=p,p-1,\dots,\ {1\over 2}.\nonumber
\end{eqnarray}

Let us add a few precisions concerning the UIR's which extend to
the conformal group SO$_0(2,4)$. First recall that, in our view,
these UIR's will  correspond to the massless fields in de Sitter
space. Masslessness will in fact be synonymous of conformal
invariance throughout this paper. In Ref. \cite{barutbohm}, the
reduction of the SO$_0(2,4)$ unitary irreducible representations
to the de Sitter subgroup SO$_0(1,4)$ UIR's are examined. It is
found that the SO$_0(1,4)$ UIR's which can be extended to UIR's of
the conformal group are the following:
\begin{itemize}
\item [-] The scalar representation with $p=0$, $q=1$ and $\langle
Q^{(1)}\rangle=2$, which, in the above classification, belongs to
the {\bf complementary series} of UIR. In that case, the
SO$_0(2,4)$ representation remains irreducible when restricted to
the SO$_0(1,4)$ subgroup. \item [-] The UIR's characterized by
$p=q=\frac{1}{2},1,\frac{3}{2},2,..$, which correspond to some
terms of the {\bf discrete series} of UIR. For any values such
that $p=q$, there are two inequivalent unitary irreducible
representations of SO$_0(2,4)$ and both remain irreducible when
restricted to SO$_0(1,4)$. These two UIR's denoted
$\Pi^{\pm}_{p,p}$  differ in the sign of the parameter $k_0=\pm p$
connected to a subgroup SO$(3)$ and  there is no operator in
SO$_0(2,4)$ which changes the value of that sign.    Therefore
these two UIR's are distinguished by an entity which we are
allowed to name the helicity.

\end{itemize}

We have pictured  these representations (up to $p=3$) in terms of
$p$ and $q$ in Figure 3. The symbols ${\color{blue}\bigcirc}$ and
$\square$ stand for the discrete series with half-integer and
integer values of $p$ respectively. Note that except when $q=0$,
each symbol $\bigcirc$ or $\square$ actually represents  two UIR's
due to the $\pm$ sign in $\Pi^{\pm}_{p,q}$.  The complementary
series is represented in the same diagram  by bold lines. The
principal series is represented in the Re$(q)=1/2$ plane by dashed
lines. We have superposed  the three discrete series of
representation with values $p=1/2,3/2,5/2$ , Re$(q)=1/2$ and
Im$(q)=0$ to the principal series in order to show how these two
diagrams fit together. Note that the substitution $p\rightarrow
p,\, q\rightarrow (1-q)$ or  $p\rightarrow (q-1),\, q\rightarrow
(p+1)$ do not alter the eigenvalues of the Casimir operators; the
representations (unitary or not) with labels $\Delta=(p,q)$,
$\Delta=(p,1-q)$ and $\Delta=(q-1,p+1)$ are  said to be ``Weyl
equivalent''. Weyl equivalent points can be localized in figure 3.
For instance  start from the points $q=\frac{1}{2}$ and
$p=0,1,2,\dots$: the bold lines (complementary series here) on the
right hand side of these points are Weyl equivalent to the bold
lines on the left hand side, including the limiting points
belonging to the discrete series in the case $p>0$.

\begin{figure}
\begin{center}
\input{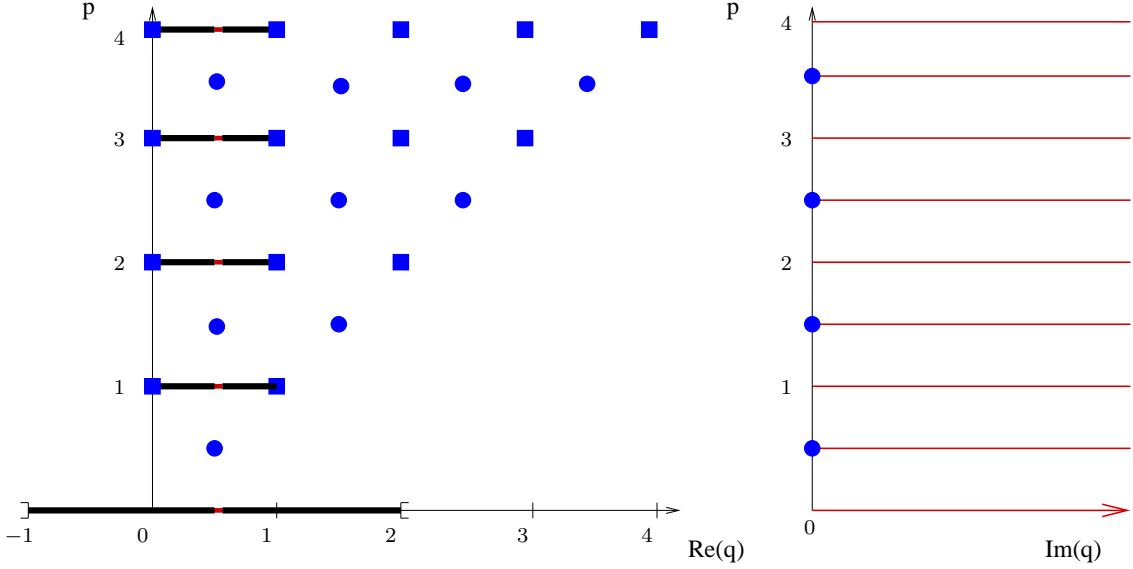}
\end{center}
\label{f1}
\caption{Set of UIR of the dS group.}
\end{figure}

\section{Gupta-Bleuler triplet of the electromagnetic field}\label{appB}

Let us briefly present the Gupta-Bleuler triplet  for the
electromagnetic field. We actually would like to give a hint of how
the indecomposable representation structure of the Poincar\'{e}
group is implemented starting with the usual plane waves.  The
electromagnetic field $A_{\mu}(x)$, defined on the 4-dimensionnal
Minkowski space-time $M$ with metric tensor
$\eta_{\mu\nu}=(+1,-1,-1,-1)$, satisfies the Maxwell equation
$$\Box A_{\mu}(x) -\partial_{\mu}(\partial^{\nu}A_{\nu}(x))=0\qquad \mbox{with}\qquad\Box=\eta^{\mu\nu}\partial_{\mu}
\partial_{\nu}\qquad \mbox{and}\qquad x=(t,{\bf{x}})\,.$$
Now the four-vector potential $A_{\mu}(x)$ is defined up to a gauge
transformation
$$A_{\mu}(x)\longrightarrow A_{\mu}'(x)=A_{\mu}(x)+\partial_{\mu}\Lambda(x)\;.$$
This gauge  symmetry couples the components of the vector potential
and reduces the effective degrees of freedom from 4 to 2. It is left
free to quantize only these two independent (physical) components of
the field (Coulomb gauge). Since the vector potential is covariantly
described by a four-vector, we assure manifest covariance using the
Lorentz gauge, characterized by the condition
$$\partial_{\mu}\,A^{\mu}(x)=0\,.$$
Independently of any quantization scheme, we are now in position to
define the Gupta-Bleuler triplet $V_g \subset V \subset V'$ carrying
the indecomposable structure of the related unitary irreducible
representation of the Poincar\'{e} group.
\begin{itemize}
\item[-]
The space  $V'$ is the space  of all solutions of the field equation
including negative norm  solutions.
\item[-] It contains a closed subspace $V$ of solutions
satisfying the Lorentz condition. The invariant subspace $V$ is not
invariantly complemented in $V'$.
\item[-]The subspace $V_g$ of $V$
consists of all positive energy gauge  solutions of the form
$A_{\mu}=\partial_{\mu}\Lambda$. These are orthogonal to every
element in $ V$ including themselves. They form an invariant
subspace of   $V$ but admit no  invariant complement in $V$.
\end{itemize}
 The usual Klein-Gordon inner product is
indefinite in $V'$, semi-definite in $V$  and gets
positive-definiteness in the quotient space $V/V_{g}$. The latter is
the physical state space.

The Poincar\'e group acts on the physical  (or transverse) space
$V/V_{g}$ through the massless, helicity $\pm 1$ unitary
representation $\mathcal{P}(0,1)\oplus \mathcal{P}(0,-1)$
\cite{bertrand}. Now since $V_{g}$  and $V'/V$ (scalar states) carry
a representation equivalent to $\mathcal{P}(0,0)$ (massless scalar
UIR of the Poincar\'e group) it is possible to write the
 representation on positive energy states as
$$\underbrace{\mathcal{P}(0,0)}_{\mbox{scalar states}}
\longrightarrow\;\underbrace{ \mathcal{P}(0,1)\oplus
\mathcal{P}(0,-1)}_{\mbox{physical states}}\longrightarrow
\underbrace{ \mathcal{P}(0,0)}_{\mbox{gauge states}}\,,$$ where the
arrows indicates the leak under the group action. We now analyse
this indecomposable structure in terms of the (vector) plane waves
with components:
$$\phi_{\mu}^{r}(x)= {\epsilon}_{\mu}^{r}(k)\frac{{e^{\displaystyle
i({\vec{k}}\cdot{\vec{x}}-\omega_{{\vec{k}}}t)}}}{\sqrt{2\omega_{\vec
k }}} \qquad\mbox{with}\quad
k_0=\omega_{{\vec{k}}}=\sqrt{\vert{\vec{k}}\vert^{2}} \,,$$ where
the polarization vectors are
$$ \epsilon^{0}(k)= (\epsilon^{0}_{\mu}(k))=   \left(\begin{array} {c}
1 \\ \vec{0}  \end{array}\right),\; \epsilon^{3}(k)=  \left(\begin{array} {c} 0 \\
\vec{k}/\omega_{\vec k} \end{array}\right),$$ $$
\epsilon^{i}(k)= \left(\begin{array} {c} 0 \\
\vec{\epsilon_{i}}
\end{array}\right)\;\; \mbox{with}  \;\;\vec
k\cdot\vec{\epsilon}_{i}(k)=0,\;\;
\vec\epsilon_i(k)\cdot\vec\epsilon_j(k)=\delta_{ij}\;\;
\mbox{for}\;\; i=1,2.$$ These vectors obey the following identity:
$$
\quad{\epsilon}_{\mu}^{r}(k)\,{{\epsilon}}_{\nu}^{r'}(k)\,
\eta^{\mu\nu}=\eta^{rr'}.
$$

As usual, the polarization $\epsilon^i(k)$ with $i=1,2$ are the
physical transverse polarizations. The plane waves are normalized
(in the Bohr integral sense) to
$$\vert\vert\phi^{r}\vert\vert^2=+1 \quad\mbox{for} \quad r=1,2,3
\quad\mbox{and} \quad \vert\vert\phi^{0}\vert\vert^2=-1.
$$
The gauge states $\phi_g\in V_g$ and the scalar states $\phi_s\in
V'/V$ can be written
\begin{equation*}
\phi^g_{\mu}=\frac{1}{\sqrt 2
}\left(\phi^0_{\mu}+\phi^3_{\mu}\right)=\frac{k_{\mu}e^{ik\cdot
x}}{2\omega_{\vec k}^{3/2}}\,,\qquad \phi^s_{\mu}=\frac{1}{\sqrt 2
}\left(\phi^0_{\mu}-\phi^3_{\mu}\right)=\frac{\tilde{{k}}_{\mu}e^{ik\cdot
x}}{{2\omega_{\vec k}^{3/2}}}\qquad \tilde{{k}}_{0}=k_{0},\;
\tilde{{k}}_{i}=-k_{i}\;,
\end{equation*}
and satisfy
$$\langle
\phi^s,\phi^g \rangle=1,\qquad\vert\vert\phi^{g}\vert\vert^2=0\,,
\qquad\vert\vert\phi^{s}\vert\vert^2=0 \,.$$ A general solution of
the field equation can be written
$$\phi(x)=a_g\phi^{g}+a_{1}\phi^{1}+a_{2}\phi^{2}+a_s\phi^{s}\,.$$
We now consider the group action  in order to display the
indecomposable structure.  In terms of vector components, the
Poincar\'{e} group acts as
$$(U(a,\Lambda)\phi)_{\mu}(x)=\Lambda_{\mu}^{\nu}\phi_{\nu}(\Lambda^{-1} (x-a))\,.$$
First of all let us show that starting with a physical state, the
group action will yield transverse states and gauge states. For this
we choose $(k_{\mu})=(1,1,0,0)$ and the transverse state
$(\phi_{\mu}^{1})$ with $(\epsilon^{1}_{\mu}(k))=(0,0,1,0)$. Let us
consider the group action $U(0,\Lambda)$
\begin{equation*}
\mbox{with} \quad \Lambda=\left(
\begin{array}{cccc}
\cosh\theta&0&\sinh\theta&0\\
0&1&0&0\\
\sinh\theta&0&\cosh\theta&0\\
0&0&0&1
\end{array} \right)
\quad\mbox{which yields} \quad \Lambda k=\left(\begin{array} {c} \cosh\theta \\
1\\ \sinh\theta\\0
\end{array}\right)\,,
\quad \Lambda \epsilon^{1}(k)=\left(\begin{array} {c} \sinh\theta \\
0\\ \cosh\theta\\0
\end{array}\right)\,.
\end{equation*}
One gets
$$(U(0,\Lambda)\phi^{1})(x) =\Lambda
\epsilon^1(k)e^{i\Lambda k\cdot x}=\tanh\theta\left[\left(\begin{array} {c} \cosh\theta \\
1\\ \sinh\theta\\0
\end{array}\right)- \left(\begin{array} {c} 0 \\
1\\ -1/\sinh\theta\\0
\end{array}\right)\right] e^{i\Lambda k\cdot x}
$$
and therefore
$$(U(0,\Lambda)\phi^{1})(x)=
\alpha\underbrace{(\Lambda k) \;\;e^{i\Lambda k\cdot
x}}_{\mbox{gauge state}} +\beta \underbrace{\epsilon^1(\Lambda
k)e^{i\Lambda k\cdot x}}_{\mbox{transverse state}}\;\;_,$$ where
$\epsilon^1(\Lambda k)$ is transverse with respect to $\Lambda k$.

Let us now start with a scalar state in order to see that the group
action will generate scalar states, transverse states as well as
gauge states. With the same vector $(k_{\mu})=(1,1,0,0)$ we define
the scalar state $(\phi_{\mu}^s)$ with
$(\epsilon^{s}_{\mu}(k))=(1,-1,0,0)$. We again consider  the group
action $U(0,\Lambda)$
\begin{equation*}
\mbox{with} \quad \Lambda=\left(
\begin{array}{cccc}
\cosh\theta&0&\sinh\theta&0\\
0&1&0&0\\
\sinh\theta&0&\cosh\theta&0\\
0&0&0&1
\end{array} \right)
\quad\mbox{which yields} \quad \Lambda k=\left(\begin{array} {c} \cosh\theta \\
1\\ \sinh\theta\\0
\end{array}\right)\,,
\quad \Lambda \epsilon^{s}(k)=\left(\begin{array} {c} \cosh\theta \\
-1\\ \sinh\theta\\0
\end{array}\right)\,.
\end{equation*}
One gets
$$(U(0,\Lambda)\phi^{s})(x)=\Lambda
\epsilon^s(k)e^{i\Lambda k\cdot x}=\left[\alpha\left(\begin{array} {c} \cosh\theta \\
-1\\ -\sinh\theta\\0
\end{array}\right)+\beta \left(\begin{array} {c} \cosh\theta \\
1\\ \sinh\theta\\0
\end{array}\right)+\gamma \left(\begin{array} {c} 0 \\
\tanh\theta\\ -1/\cosh\theta\\0
\end{array}\right)\right] e^{i\Lambda k\cdot x}
$$
and therefore
$$(U(0,\Lambda)\phi^{s})(x)=\alpha
\underbrace{\epsilon^s(\Lambda k)e^{i\Lambda k\cdot
x}}_{\mbox{\footnotesize scalar state}} +\beta\underbrace{(\Lambda
k) e^{i\Lambda k\cdot x}}_{\mbox{\footnotesize gauge state}}
+\gamma{}\underbrace{\epsilon^i(\Lambda k)e^{i\Lambda k\cdot
x}}_{\mbox{\footnotesize transverse state}}\;\;_,$$ where
$\epsilon^i(\Lambda k)$ with $i=1,2$ is transverse with respect to
$\Lambda k$ and with
$\alpha=1/(\cosh^2\theta),\quad\beta=\tanh^2\theta,\quad\gamma=-2\tanh\theta$.

\section{The two-point function from maximally symmetric bivectors in ambient space}\label{appC}

Following Allen and Jacobson in reference \cite{allen1} we
express here the two-point functions in de Sitter space (maximally
symmetric) in terms of bivectors.

Maximally symmetric bivectors are functions of two
points $(x,x')$ which behave like vectors under coordinate
transformations at either point. The bivectors are called
maximally symmetric if they respect the de Sitter invariance.

As shown in reference \cite{allen1}, any maximally symmetric
bivector can be expressed as a sum of products of three  basic
tensors. The coefficients in this expansion are functions of the
geodesic distance $\mu(x,x')$, that is the distance along the
geodesic connecting the points $x$ and $x'$ (note that $\mu(x,x')$
can be defined by unique analytic extension also when no geodesic
connects $x$ and $x'$). In this sense,  these fundamental tensors
form a complete set.  They can be obtained by differentiating the
geodesic distance:
$$n_{a}=\nabla_{a}\mu(x,x'),\qquad n_{a'}=\nabla_{a'}\mu(x,x')$$
and through the parallel propagator
$$ g_{ab'}=-c^{-1}(\z)\nabla_{a} n_{b'}+n_{a}n_{b'}\;.$$
The geodesic distance is implicitly defined \cite{brmo} for
$\z=-H^{2}x\cdot x'$ by
\begin{eqnarray*}
\z&=&\cosh{(\mu H)}\quad\mbox{for $x$ and $x'$ timelike separated,}
\\
 \z&=&\cos{(\mu H)}\quad\mbox{for $x$ and $x'$ spacelike separated such
 that}\quad \vert x\cdot x'\vert <H^{-2}.
\end{eqnarray*}
The two-point function in terms of the basis bi-vectors reads
$$
T_{ab'}(x,x')=\alpha(\mu) g_{ab'}+\beta(\mu) n_{a} n_{b'}\,.
$$
Since in this paper  we  work in ambient space, let us accordingly
re-express the basic bi-vectors in terms of the corresponding
notations.

{\prop
In ambient space notations ($x \in \setR^5$ and the constraint
$x\cdot x=-H^{-2}$), the basic bi-vectors corresponding to $n_a$,
$n_{a'}$ and $ g_{ab'}$ can be chosen as
$$\bc\mu(x,x'),\qquad\bpc\mu(x,x'),\qquad \theta_{\alpha}\cdot\theta'_{\beta'}.$$}

\DEM

One merely has to consider the restriction to the hyperboloid given by
$$T_{ab'}(x,x')=\ab T_{\alpha\beta'}\;.$$

$\bullet$ When $\z=\cos(\mu H)$, one finds
$$n_{a}=\a\bar\partial_{\alpha}\mu(x,x')=\a{{H(\theta_{\alpha}\cdot
x')}\over{\sqrt{1-\z^2}}},\quad\qquad
n_{b'}=\bb\bar\partial^{'}_{\beta'}\mu(x,x')=\bb{{H(\theta'_{\beta'}\cdot
x)}\over{\sqrt{1-\z^2}}}\;,$$ and
$$\nabla_{a}n_{b'}=\ab
\bar\partial_{\alpha}\bar\partial'_{\beta'}\mu(x,x')=c(\z)\left[\z
n_{a}n_{b'}- \ab\theta_{\alpha}\cdot\theta'_{\beta'} \right] \,,$$
with $\displaystyle{c(\z)=-\frac{H}{\sqrt{1-\z^{2}}}}$.

$\bullet$
When $\z=\cosh(\mu H)$, $ n_{a}$, $ n_{b'}$   are multiplied by $i$ and
$c(\z)$ becomes $-\frac{iH}{\sqrt{1-\z^{2}}}$.

In both cases we
have
$$
\ab\theta_{\alpha}\cdot\theta'_{\beta'}
=g_{ab'}+(\z-1)n_{a}n_{b'}\;.
$$
\QED

\section{The two-point function from the field modes}\label{appE}
\setcounter{equation}{0}
We have shown that the  vector two-point function can be expressed
as

\begin{equation} \label{eq:tpf4}
W_{\alpha \alpha'}(z,z') = c_0 \int_{\gamma}\left(
\theta_{\alpha}\cdot\theta'_{\alpha'}-\frac{2\,\bar\xi'_{\alpha'}\bar\xi_{\alpha}}{(H
z\cdot\xi)^2}\right)\phi_1(z)\phi_2(z') d\sigma_{\gamma}(\xi)\,,
\end{equation}

where $\phi_1(z)=(H z\cdot \xi)^{-1}$ and $\phi_2(z)=(H z\cdot
\xi)^{-2}$. This has been done starting with the formula
$$
W_{\alpha \alpha'}(z,z')=-H^{-2}\bar \partial_{\alpha}\,
\theta'_{\alpha'}\cdot\bar
\partial\, W_0(z,z')-x\cdot\theta'_{\alpha'} \bar
\partial_{\alpha}W_0(z,z')\,,
$$
based on the general maximally symmetric bivector form. This
two-point function is also equal to $$ W_{\alpha
\alpha'}(z,z')=-H^{-2}\bar \partial'_{\alpha'}\,
\theta_{\alpha}\cdot\bar
\partial'\, W_0(z,z')-x'\cdot\theta_{\alpha} \bar
\partial'_{\alpha'}W_0(z,z')\,,
$$ which  yields
$$
W_{\alpha \alpha'}(z,z') = 2 \theta_{\alpha}\cdot\theta'_{\alpha'}
W_0(z,z')+\theta\cdot x'\bar\partial'W_0(z,z')-c_0
\int_{\gamma}6\frac{\,\bar\xi'_{\alpha'}\bar\xi_{\alpha}}{(H
z'\cdot\xi)^2}\phi_1(z)\phi_2(z') d\sigma_{\gamma}(\xi)\,.
$$
Thus it is established that

\begin{eqnarray}\label{eq:prop}
-c_0 \int_{\gamma}6\frac{\,\bar\xi'_{\alpha'}\bar\xi_{\alpha}}{(H
z'\cdot\xi)^2}\phi_1(z)\phi_2(z') d\sigma_{\gamma}(\xi)&=& -
\theta_{\alpha}\cdot\theta'_{\alpha'} W_0(z,z')-\theta\cdot
x'\bar\partial'W_0(z,z') \nonumber\\
&-&c_0\int_{\gamma}2\frac{\,\bar\xi_{\alpha'}\bar\xi_{\alpha}}{(H
z\cdot\xi)^2}\phi_1(z)\phi_2(z') d\sigma_{\gamma}(\xi)\,.
\end{eqnarray}

We are now in position to calculate the two-point function with the
help of the field modes.
$$
W_{\alpha \alpha'}(z,z') =c_0\int_{\gamma}
\eta_{\lambda\lambda'}\left( \bar Z^{\lambda}_{\alpha}- 2
\frac{Z^{\lambda}\cdot \xi}{(H x\cdot\xi)^2}\right)$$ $$\left(
2\bar Z_{\alpha'}^{'\lambda'}- 6 \frac{Z^{\lambda'}\cdot \xi}{(H
x'\cdot\xi)^2} \bar \xi'_{\alpha'}-2\frac{Z^{\lambda'}\cdot
x'}{x'\cdot \xi} \,\bar
\xi'_{\alpha'}\right)\phi_1(z)\phi_2(z')d\sigma_{\gamma}(\xi) \,.
$$
For $\eta_{\lambda\lambda'}Z_{\alpha}^{\lambda}
Z_{\beta}^{\lambda'}=\eta_{\alpha\beta}$ and using $\xi^2=0$ one
gets
$$
W_{\alpha \alpha'}(z,z') =2\theta_{\alpha}\cdot\theta'_{\alpha'}
W_0(z,z')+\theta\cdot x'\bar\partial'W_0(z,z') -c_0
\int_{\gamma}6\frac{\,\bar\xi'_{\alpha'}\bar\xi_{\alpha}}{(H
z'\cdot\xi)^2}\phi_1(z)\phi_2(z') d\sigma_{\gamma}(\xi) \,.
$$
Now with the help of Eq. (\ref{eq:prop}) one finally finds that
this expression coincides with the expression (\ref{eq:tpf4}).

\end{document}